\documentclass{neuthist18}

\def\be{\begin{equation}}
\def\ee{\end{equation}}
\def\bea{\begin{eqnarray}}
\def\eea{\end{eqnarray}}

\newcommand{\Photo}{\includegraphics[height=35mm]{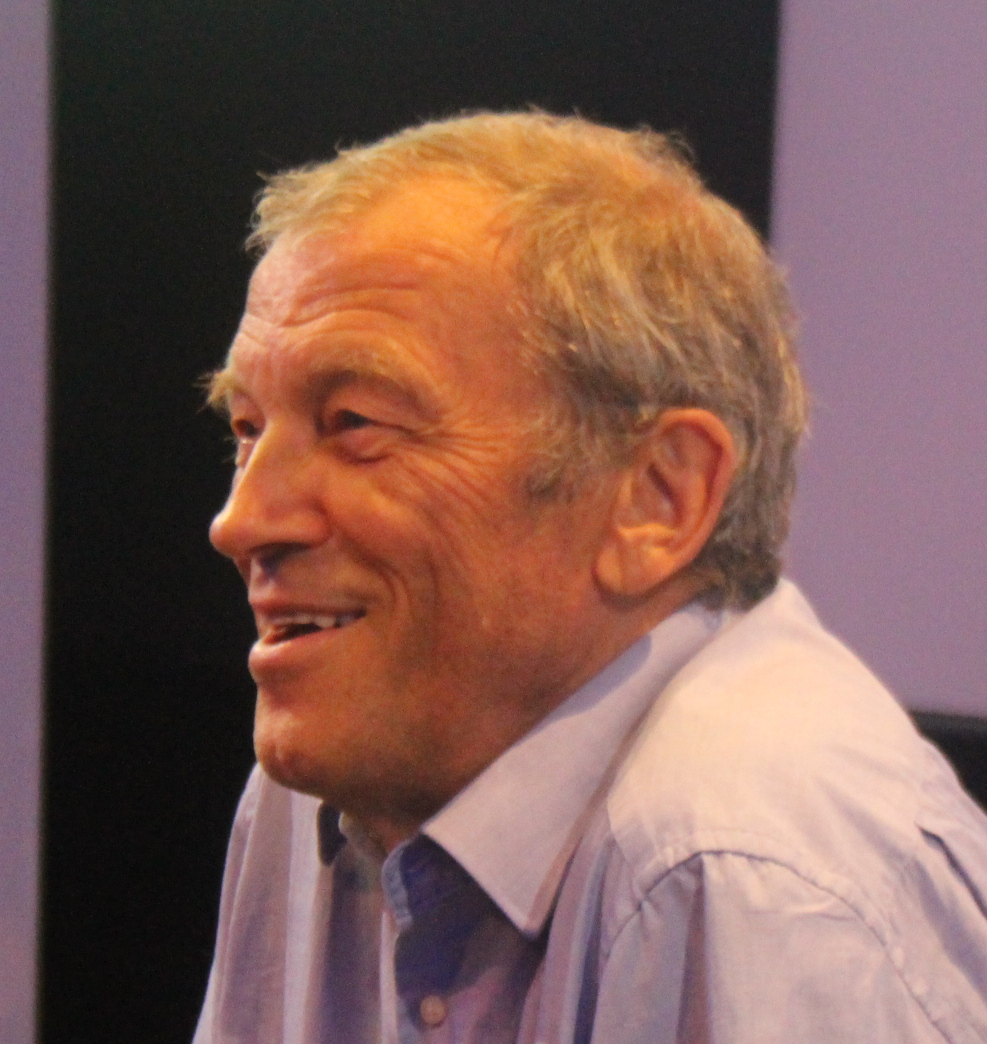}}

\begin{document}
\vspace*{4cm}
\title{History of high-energy neutrino astronomy}

\author{C.\,Spiering}

\address{DESY, Platanenallee 6, D-15738 Zeuthen, Germany}

\maketitle\abstracts{
This talk sketches the main milestones of the path towards 
cubic kilometer neutrino telescopes. It starts with the first conceptual ideas in the late 1950s and describes the emergence of concepts for detectors with a realistic discovery potential 
in the 1970s and 1980s. After the pioneering project DUMAND close to Hawaii was terminated in 1995, the further development was carried by NT200 in Lake Baikal, AMANDA at the South Pole and ANTARES in the Mediterranean Sea.  In 2013, more than half a century after the first concepts, IceCube has discovered extraterrestrial high-energy neutrinos and opened a new observational window to the cosmos -- marking
a milestone along a journey which is far from being finished.
}

\section{From first concepts to the detection of atmospheric neutrinos}
\label{sec-concepts}

The initial idea of neutrino astronomy beyond the solar system  
rested on two arguments: The first was the
expectation that a supernova stellar collapse in our galaxy would be accompanied
by an enormous burst of neutrinos in the 5-10 MeV range. The second was the expectation 
that fast rotating pulsars must accelerate charged particles in their Tera-Gauss magnetic
fields. Either in the source or on their way to Earth they must hit matter, generate
pions and neutrinos as decay products of the pions.

The first ideas to detect cosmic high energy neutrinos underground or underwater
date back to the late fifties (see \cite{Spiering-EPJ} for a detailed
history of cosmic neutrino detectors).  In the 1960 Annual Review of Nuclear Science,
K.\,Greisen and F.\,Reines discuss the motivations and prospects for such detectors.
In his paper entitled \textsc{Cosmic Ray Showers} \cite{Greisen-1960}, Greisen writes:
\textsl {``As a detector,
we propose a large Cherenkov counter, about 15 m in diameter, located
in a mine far underground. The counter should be surrounded with photomultipliers
to detect the events, and enclosed in a shell of scintillating material to distinguish
neutrino events from those caused by $\mu$ mesons. Such a detector would be
rather expensive, but not as much as modern accelerators and large
radio telescopes. The mass of the sensitive detector could be about 3000 tons of 
inexpensive liquid.''}
Later he estimates the rate of neutrino events from the Crab Nebula
as one count per three years and
optimistically concludes:
\textsl{``Fanciful though this proposal seems, we suspect that within the next decade
cosmic ray neutrino detection will become one of the tools of both
physics and astronomy.''}

F.\,Reines in his article \textsc{Neutrino Interactions} \cite{Reines-1960} is more conservative
with respect to extraterrestrial neutrinos: 
\textsl{``At present no acceptable theory of the origin and extraterrestrial diffusion
exists so that the cosmic neutrino flux can not be usefully predicted.''}
At this time, he could not be aware of the physics potential of atmospheric
neutrinos and continues:  
\textsl{``The situation
is somewhat simpler in the case of cosmic-ray neutrinos} ("atmospheric neutrinos"
in present language. C.S.) \textsl{ -- they are both more predictable and of less
intrinsic interest.''}

In the same year, on the 1960 Rochester Conference, M.\,Markov published his
ground breaking idea   \cite{Markov-1960}
\textsl{``...to install detectors deep in a lake or a sea and  to determine the 
direction of  charged particles with the help of Cherenkov radiation. ....We consider $\mu$ mesons
produced in the ground layers under the detector''.}
This appeared to be the only way to reach detector volumes beyond the scale of
$10^4$ tons.

During the sixties, no predictions or serious estimates for neutrino fluxes from 
specific cosmic accelerators were 
published \footnote{The only quantitative flux estimate was related to a {\it diffuse} flux of 
cosmic neutrinos due to interactions of cosmic rays with the 3K microwave background radiation
\cite{BZ}.}. Actually, many of the objects nowadays considered
as top candidates for neutrino emission were discovered only in the sixties and
seventies (the first quasar 1963, pulsars 1967, X-ray binaries with a black hole
1972, gamma ray bursts 1973).  The situation changed dramatically in the seventies,
when these objects were identified as possible neutrino emitters, triggering
an enormous amount of theoretical activity. 

Different to extraterrestrial neutrino fluxes, the calculation of the flux of atmospheric 
neutrinos became more reliable. 
First serious estimates were published in the early sixties
\cite{Zatsepin-1961,Cowsik-1963,Osborne-1965}. 
These pioneering attempts are described in
detail in the recollections of Igor Zheleznykh \cite{Zheleznykh-2008} 
who was a student of Markov. In his diploma work from 1958 he  
performed early estimates for the flux of atmospheric neutrinos
and for the flux of neutrinos from the Crab nebula.
The real explosion of papers on atmospheric neutrinos, however, happened between 1980 and 1990
when the large underground detectors 
became operational and the field turned into a
precision science (see the talks of John Learned and Paolo Lipari at this conference).

In contrast to investigating atmospheric and supernova neutrinos,
the study of high-energy extraterrestrial neutrinos had the inherent risk that
no reliable predictions for the expected fluxes could be made. Under these circumstances
it appeared logical to tackle this problem with the largest devices conceivable,
with underwater detectors of the kind which Markov had proposed in 1960.

\section{DUMAND}
\label{sec-dumand}

The history of underwater neutrino telescopes starts with a project which
eventually was terminated but left an incredibly rich legacy of ideas and technical
principles: The DUMAND project. DUMAND stands for Deep Underwater Muon and
Neutrino Detector. Its early history is excellently covered in a \textsc{Personal
history of the Dumand project} by A.\,Roberts \cite{DUMAND-Roberts}. 

At the 1973 International Cosmic Ray Conference (ICRC), a few physicists
including F.\,Reines and J.\,Learned (USA),
G.\,Zatsepin (USSR) and S.\,Miyake (Japan)
discussed a deep-water detector to clarify puzzles in muon depth-intensity
curves. The puzzles faded away, but it was obvious that such a
detector could also work for neutrinos. 

The year 1975 saw the first of a -- meanwhile legendary -- series of DUMAND 
Workshops, this one at Washington State University \cite{DUMAND-1975}. A survey of possible
sites converged on the Pacific Ocean close to Hawaii, since it offered
deep locations close to shore. A year later, a two-week workshop took
place in Honolulu \cite{DUMAND-1976}.
At that time, three options for a deep sea array were discussed:

\begin{itemize}
\item UNDINE (for "UNderwater Detection of Interstellar Neutrino Emission") was 
intended to detect neutrinos from supernova collapses from far beyond our
own Galaxy (leaving the Galactic Supernovae to underground detectors).
Based on overoptimistic assumptions of the neutrino energy spectrum,
it was soon discarded.
\item ATHENE (for "ATmospheric High-Energy Neutrino Experiment") was tailored
to high-energy particle physics with atmospheric neutrinos. 
\item UNICORN (for "UNderwater Interstellar COsmic-Ray Neutrinos") had the
primary goal to search for high-energy extraterrestrial neutrinos.
\end{itemize}

At the 1976 workshop and, finally, at the 1978 DUMAND workshop \cite{DUMAND-1978} 
the issue was settled in favour of an array
which combined the last two options, ATHENE and UNICORN. 

The principle of the detector was -- as already suggested by Markov --
to record upward-travelling muons generated in 
charged current muon neutrino interactions. 
The upward signature guarantees the neutrino origin of the muon. 
Since neutrino oscillations were not
considered at that time, a the flavor ratio at Earth was assumed to be the same
as in the source ($\nu_e:\nu_{\mu}:\nu_{\tau} = 1:2:0$ for 
$\pi \rightarrow \mu \rightarrow e$ decay) suggesting to focus on muon neutrino detection.

\begin{figure}[ht]
\begin{center}
\includegraphics[width=4.5cm]{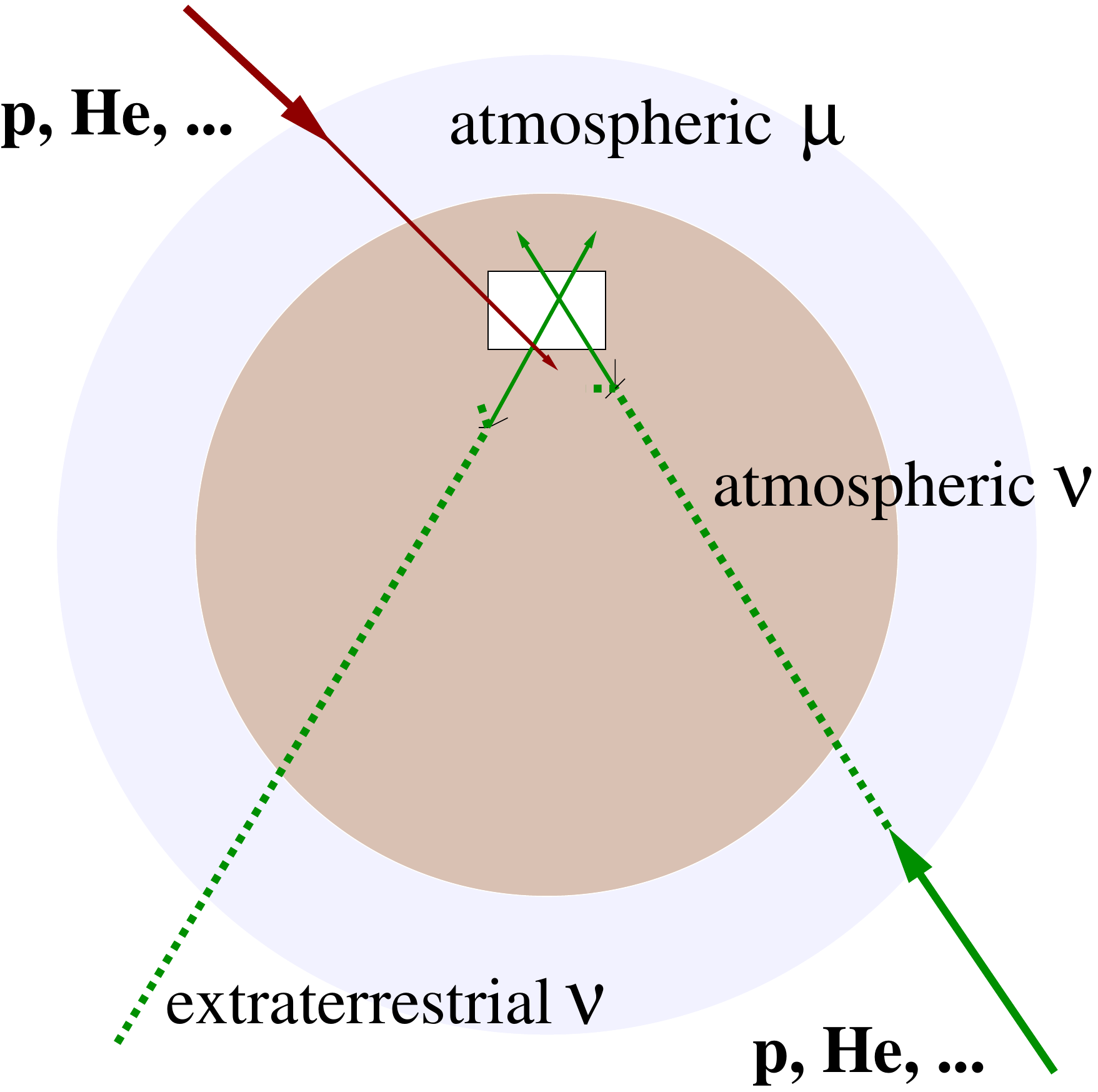}
\hspace{2cm}
\includegraphics[width=5.5cm]{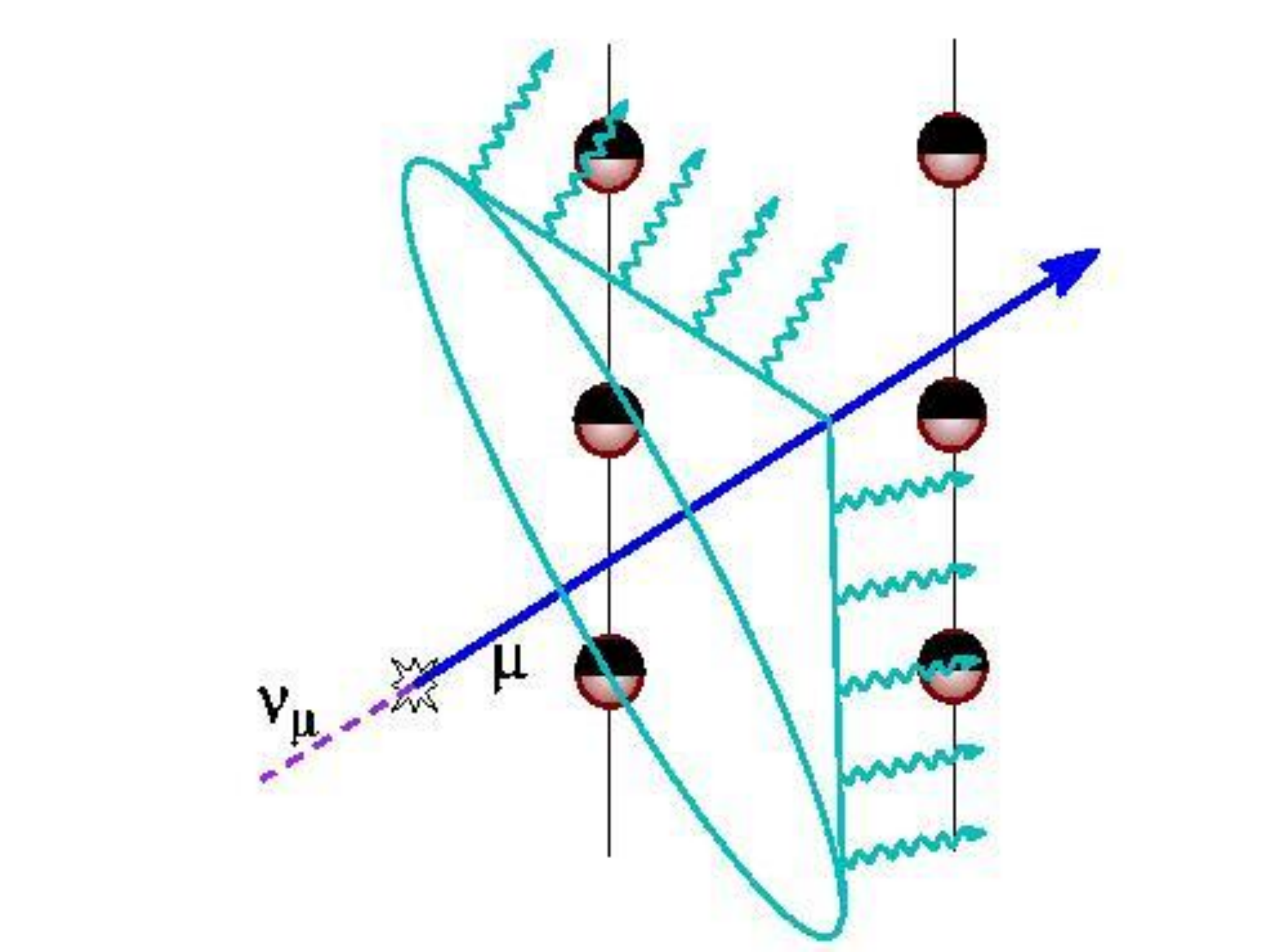}
\end{center}
\caption{
{\it Left:} Sources of muons in deep underwater/ice detectors. Cosmic nuclei -- protons (p),
$\alpha$ particles (He), etc.\ -- interact in the Earth atmosphere
(light-colored). Sufficiently energetic muons produced in these interactions
("atmospheric muons'') can reach the detector (white box) from above. 
Upward-going muons must have been produced in neutrino interactions.
{\it Right:} Detection principle for muon tracks.  
\label{sources}
}
\end{figure}

Since the sixties, a large depth was recognized as necessary in order to suppress
downward-moving muons which may be mis-reconstructed as upward-moving ones
(Fig.\,\ref{sources}, left). Apart from these, only one irreducible background to
upward moving extra-terrestrial neutrinos remains: atmospheric neutrinos.
This background cannot be reduced by going deeper. On the other hand, it provides a
standard calibration source and a reliable proof of principle.

\begin{figure}[ht]
\begin{center}
\includegraphics[width=13.5cm] {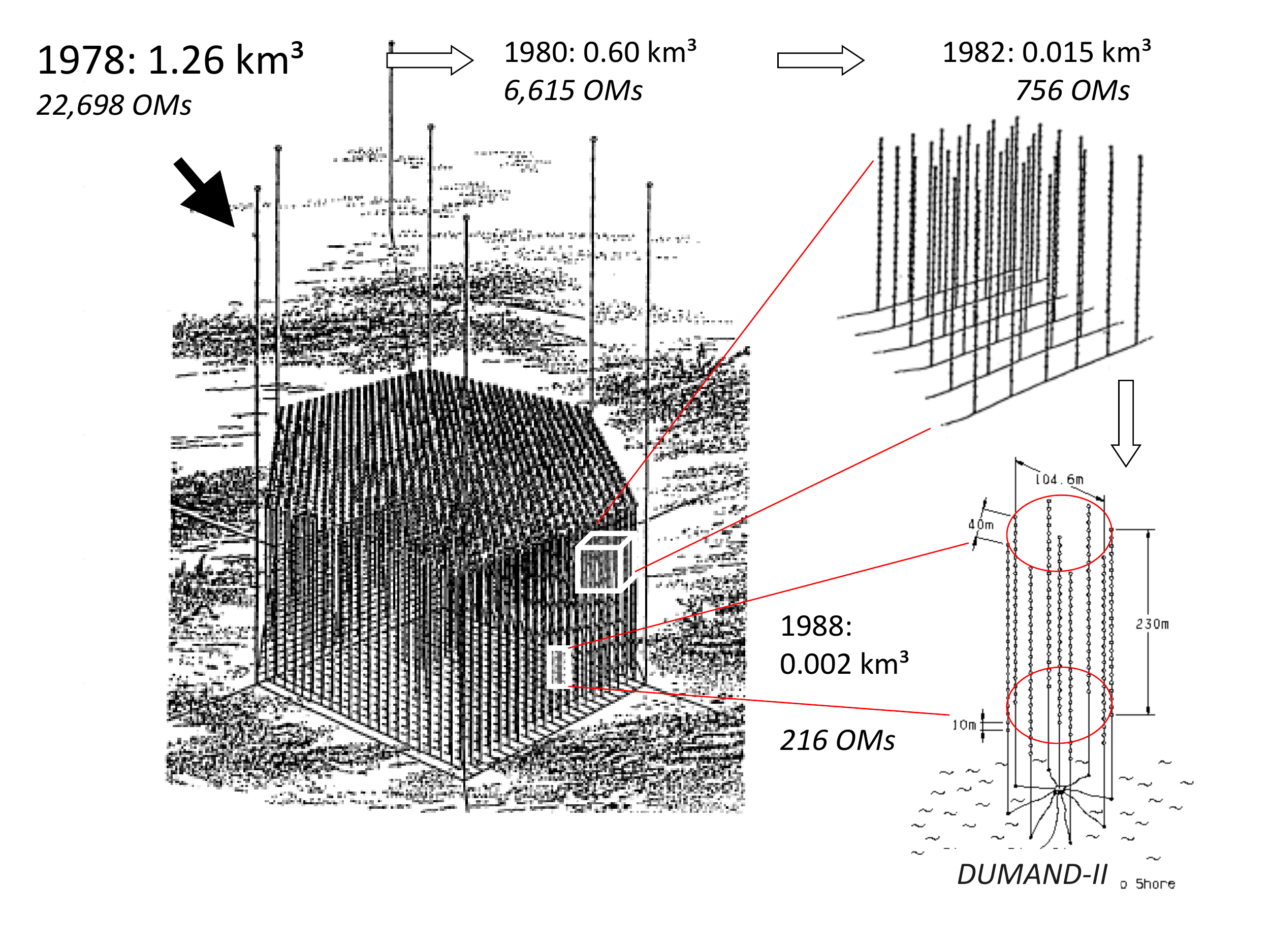}
\end{center}
\vspace{-7mm}
\caption{   
The originally conceived DUMAND cubic kilometer detector and the
phased downgrading to the 1988 plan for a first-generation underwater neutrino telescope
DUMAND-II.}   
\label{DUMAND}
\end{figure}

The DUMAND design envisaged an array of photomultiplier tubes
(PMTs) housed in glass spheres spread over 
a cubic kilometer (see Fig.\ref{DUMAND}, left). The PMTs would record arrival time and
amplitude of Cherenkov light emitted by muons or particle cascades. 
The spheres were to be attached to strings moored at the ground and held
vertically by buoys. 
From the arrival times, the direction of the muon track can be reconstructed,
and it turned out that a directional accuracy of 1\,degree is achievable. This
is of a similar size as the kinematic smearing between neutrino and muon direction 
and allows for {\it neutrino tracing}, i.e. for neutrino astronomy.

Naturally, the idea to construct a cubic-kilometer detector with more than
20\,000 large-size PMTs (see Fig.~\ref{DUMAND}) challenged
technical and financial possibilities. A.\,Roberts remembers \cite{DUMAND-Roberts}: 
\textsl{``The 1978 DUMAND
Standard Array, on closer examination, assumed more and more awesome proportions.
... 1261 sensor strings, each with 18 complex sensor modules ...
to be deployed on the ocean bottom at a depth of 5 km! The oceanographers were
amazed -- this project was larger than any other peacetime ocean project by a factor
of the order of 100. The size of the array was based on relatively scant information
on the expected neutrino intensities and it was difficult to justify in detail; the general
idea was that neutrino cross section are small and high-energy neutrinos are
scarce, so the detector had better be large.''}
Confronted with the
oceanographic and financial reality, the 1.26 km$^3$ array was abandoned. 
A half-sized configuration (1980) met the same fate, as did a much smaller array with 756 PMTs
(1982). What finally emerged as a technical project was a
216-PMT version, dubbed DUMAND-II or "The Octagon'" (eight strings at the
corners of an octagon and one in the center), 100\,m in diameter and 230\,m in
height \cite{DUMAND-Project} (see Fig.\,\ref{DUMAND}).
The plan was to deploy the detector 30\,km off the coast of Big Island, Hawaii, 
at a depth of 4.8\,km.

The evolution of the detector design, largely following
financial and technological boundary conditions, was the one side
of the story. What about the flux predictions?

At the 1978 workshop first 
investigations on neutron star binary systems as point
sources of high energy neutrinos were presented, specifically
Cygnus X-3 (D.\,Eichler/D.\,Schramm  and D.\,Helfand in 
\cite{DUMAND-1978}). The connection to the indications
for sources of TeV-$\gamma$-ray (none of them significant at that time!)
was discussed by T.\,Weekes. At the same time, the possibilities for
diffuse source detection were disfavored (R.\,Silberberg, M.\,Shapiro,
F.\,Stecker). 

The gamma-neutrino connection was discussed further by
V.\,Berezinsky at the 1979 DUMAND Workshop in Khabarovsk and Lake
Baikal \cite{DUMAND-1979}. He emphasized the concept of "hidden"
sources which are more effectively (or {\it only}) detectable by neutrinos
rather than by $\gamma$ rays. Among other mechanisms, Berezinsky also investigated the
production of neutrinos in the young, expanding shell of a supernova
which is bombarded by protons accelerated inside the shell  
("inner neutrino radiation from a SN envelope"). 
He concluded that
a 1000\,m$^2$ detector should be sufficient to detect
high-energy neutrinos from a galactic supernova over several weeks or
months after the collapse. Naturally, ten years later, in 1987, 
much attention was given to this model in the
context of SN1987. But alas! -- this supernova was at about 50 kpc distance,
more than five times farer than the Galactic center. Moreover 
all underground detectors existing in 1987 had areas much smaller
than 1000\,m$^2$. Therefore the chances to see 
inner neutrino radiation from the envelope were rather small, and actually 
"only" the MeV burst neutrinos and no high energy neutrinos have been recorded.

A large number of papers on expected neutrino
fluxes was published during the eighties. The expected neutrino  
fluxes were found to depend strongly {\it a)} on the energy spectrum of the $\gamma$-ray
sources which could only be guessed since the first uncontroversial TeV-$\gamma$ observation
was the Crab nebula in 1989 
(while all of the
earlier claims on PeV $\gamma$ rays from binary X-rays systems 
were not confirmed with more sensitive detectors) and {\it b)} on the
supposed $\nu/\gamma$ ratio which depends on the unknown thickness of matter
surrounding the source.
 
The uncertainty of expectations is reflected in Table \ref{tab-1},
taken from DUMAND-II proposal \cite{DUMAND-Project} but omitting some sources.
Pessimistic and optimistic numbers differed by 2-3 orders of
magnitude and left it open whether DUMAND-II would be able to detect
neutrino sources or whether this would remain the realm of a future cubic kilometer
array.  Two years later, V.\,Berezinsky reiterated his earlier estimates that
for neutrinos from a fresh neutron star a detector with an effective
area of 1000\,m$^2$ (i.e. a large underground detector) would be sufficient,
but that the detection of extragalactic sources would require detectors
of 0.1-1.0 km$^2$ area \cite{Berezinsky-1990}. DUMAND-II, with 25\,000 m$^2$ area, fell just
below these values. Again citing A.\,Roberts \cite{DUMAND-Roberts}:
\textsl{``These calculations 
serve to substantiate our own gut feelings. I have myself watched the progression
of steadily decreasing size ... at first with pleasure (to see it become more practical),
but later with increasing pain. ... The danger is, that if DUMAND II sees no
neutrino sources, the funding agencies will decide it has failed and, instead of
expanding it, will kill it.''}

\begin{table}
\centering
\caption{Tabulation of various gamma ray sources used for neutrino candidate
source estimates. The table is taken from the DUMAND-II proposal~\protect\cite{DUMAND-Project} but 
omitting some sources as well as information about sky position, distance, and 
assumed spectral index.}
\vspace{3mm}
\label{tab-1}       
\begin{tabular}{llllll}
\hline
  & $\gamma$-ray & $\gamma$ flux & & 
\multicolumn{2} {c} {$\mu$/yr DUMAND\,II} \\
source & energy & at Earth & luminosity & 
$\epsilon_{\nu/\gamma=1}$ & $\epsilon_{\nu/\gamma=30}$ \\
name &  (TeV) & (cm$^{-2}$ s$^{-1}$) &  (erg s$^{-1}$)& min $\gamma$ & 
max $\gamma$ \\\hline
Vela PSR & 5 & $1.8\times 10^{-12}$&$3\times 10^{32}$& 0.1 & 1506 \\
Vela X-1 & 1 & $2\times 10^{-11}  $&$2\times 10^{34}$& 0.2 & 126   \\
Crab SNR & 2 & $1.1\times 10^{-11}$&$2\times 10^{34}$& 0.2 & 438   \\
Crab PSR & 1 & $7.9\times 10^{-12}$&$6\times 10^{33}$& 0.06 & 38   \\
Geminga  & 6 & $9.5\times 10^{-12}$&$3\times 10^{33}$& 0.49 & 1506  \\
Her X-1  & 1 & $3\times 10^{-11}  $&$3\times 10^{35}$& 0.24 & 141 \\
SS443    & 1 & $<10^{-10}         $&$3\times 10^{32}$& $<0.88$& $<510$  \\
Cen X-3  & 1 & $<5.2\times 10^{-12}$&$3\times 10^{34}$&$<0.08$ & $<48$  \\
Cyg X3   & 1 & $5\times 10^{-11}  $&$3\times 10^{36}$& 0.4 & 234 \\
M31      & 1 & $2.2\times 10^{-10}$&$2\times 10^{40}$& 1.8 & 1050 \\  
Cen A    &0.3& $4.4\times 10^{-11}$&$3\times 10^{40}$& 0.14 & 6 \\
3C 273   & 5 & $<9\times 10^{-12} $&$<3\times 10^{45}$& $<0.4$ & $<1506$\\\hline
\end{tabular}
\end{table}

In 1987, the DUMAND collaboration operated a 7-PMT test string from a 
a Navy vessel \cite{DUMAND-Babson} and measured the muon intensity as a function of
depth. Carried by this success, one year later the DUMAND-II proposal was submitted to DOE and NSF.
 
DUMAND-II 
would have detected
three down going muons per minute and about 3500 atmospheric neutrinos
per year. In December 1993, a first of three prepared strings was deployed and linked to shore via a
a junction box placed on the ocean bottom and a shore cable which had
been laid some months earlier. 
However, some pressure housings developed leaks and a short circuit in the junction
box 
terminated the communication to shore.

The DUMAND progress had been slow, but had shown  
remarkable progress compared to ocean research at that time. 
This impressed oceanographers but not the
main funding organization, the Department of Energy (DOE), 
which was not used to a "try-and-try-again" mode of progress.  
Review committees without any ocean expertise judged
the project, following criteria typical for accelerator research.
Moreover, the termination of the Superconducting
Super Collider (SSC) by the US congress in 1993
created a strong risk aversion in DOE.
On the technical side, the reasons of the 1993 DUMAND 
failures had been identified and a redeployment was in preparation.
But in 1995, the mentioned circumstances regrettably
led to a termination of  the support for DUMAND.

\section{The evolution of the Baikal project}
\label{sec-baikal}


Russian participation in the DUMAND project was strong from the beginning
(represented by A.\,Chudakov, V.\,Berezinsky, L.\,Bezrukov, B.\,Dolgoshein,
A.\,Petrukhin and I.\,Zheleznykh). 
However, in the context of the Soviet
invasion in Afghanistan, in 1980 the Reagan administration terminated the cooperation.
As A.\,Roberts remembers \cite{DUMAND-Roberts}:
\textsl{``The severing of the Russian link was done with
elegance and taste. We were told, confidentially, that while we were
perfectly free to choose our collaborators as we liked, if perchance
they  included Russians it would be found that no funding was
available.''}

About the same time, however,
A.\,Chudakov proposed to use the deep water of Lake Baikal in Siberia
as the site for a "Russian DUMAND".  The advantages of Lake Baikal seemed
obvious: it is the deepest freshwater lake on Earth, with its largest depth
at nearly 1700 meter, it is famous for its clean and transparent water,
and in late Winter it is covered by a thick ice layer which allows 
installing winches and other heavy technique and deploying  
underwater equipment without any use of ships. 

In 1981, a first shallow-site experiment with small PMTs started.  
Chair of a dedicated laboratory at the Moscow Institute of Nuclear Research
(INR) became G.\,Domogatsky,
flanked by L.\,Bezrukov as leading experimentalist.
Soon a site about 30 km South-West from the outflow of Lake Baikal into
the Angara river was identified as suitable, with a distance of
3.6\,km to shore and a depth of about 1370\,m. The detector depth would be
about 1\,km.  Instruments could be deployed
in a period between late February and early April from the ice cover,  and operated
over the full year via a cable to shore.

The Russians operated several stationary strings from 1984 on,
recording downward moving muons \cite{Baikal-1984} and setting stringent
limits on the flux of magnetic monopoles 
catalyzing proton decays along their path \cite{Baikal-1986}.
At the end of the 1980s, they replaced their slow 15-cm flat photocathode PMT by the QUASAR,
a hybrid device with a 370\,mm diameter mushroom-shaped photocathode, 
with excellent 1-PE resolution,
a time jitter as small as 2\,ns and negligible sensitivity to the
Earth's magnetic field \cite{Baikal-OM-Bagduev-1999}.
In 1988, the Baikal experiment was approved as a long-term direction of research
by the Soviet Academy of Sciences and the USSR government which included
considerable funding. A full-scale detector (yet without clear definition of its size)
was planned to be built in steps of intermediate detectors of growing size.
In the same year 1988, our group from the East German Institute of High Energy Physics
in Zeuthen (part of DESY from 1992 on) joined the Baikal experiment. 

\begin{figure}[ht]
\hspace{0.8cm}
\includegraphics[width=14cm]{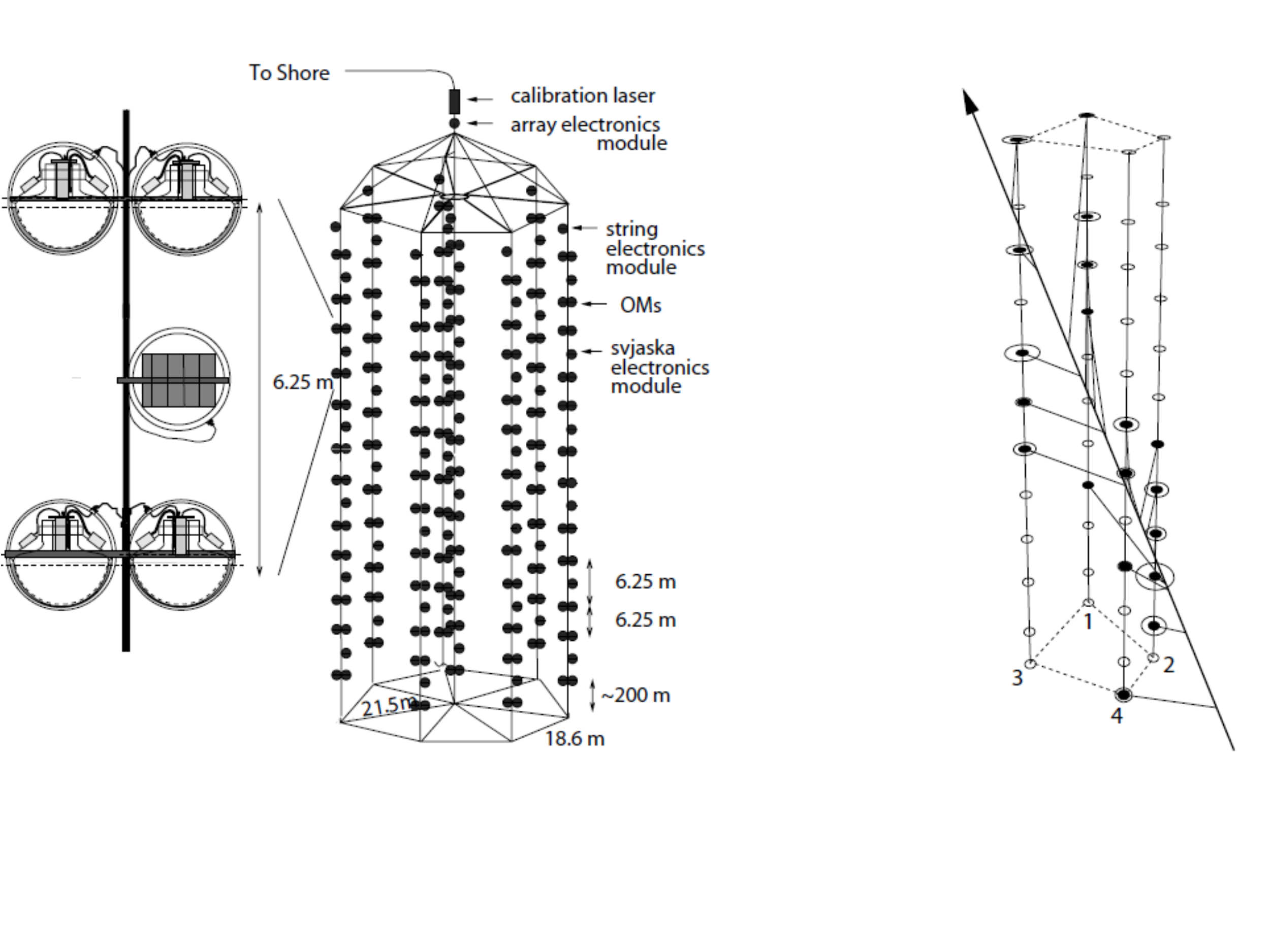}
\vspace{-2.5cm}
\caption{{\it Left:} The Baikal Neutrino Telescope NT200. 
{\it Right:} One of the first upward moving
muons from a neutrino interaction recorded with the 4-string stage of the
detector in 1996~\protect\cite{Baikal-atm-Balkanov-1999}. The Cherenkov light from the muon is recorded by 19 channels. }
\label{Baikal}
\end{figure}

From 1989 to 1992, the NT200 project was developed \cite{Baikal-Project}.
NT200 (Fig.\,\ref{Baikal},\,left) was an array of 192 optical
modules carried by eight strings which are attached to an umbrella-like
frame consisting of 7 arms, each 21.5\,m in length. 
The optical modules with the QUASAR-tubes were grouped pair-wise along a string,
with the two PMTs of a pair switched in coincidence for noise suppression.

The construction of NT200 coincided with the decay of the USSR and an
economically desperate period. Under these circumstances, 
the construction of NT200 extended over more than five years.
It started with the deployment of a 3-string array 
with 36 optical modules in March/April 1993.
With that array it was demonstrated that 3-dimensional
reconstruction of muon tracks works as expected.
The first two upward moving candidates were separated from the 1994 data.
In 1996, a 96-OM array with four NT200 strings was 
operated \cite{Baikal-atm-Balkanov-1999} and provided the first  textbook neutrinos 
like the one shown in  Fig.\,\ref{Baikal},\,right.

NT200 was completed in April 1998 and took data for about a decade. 
About 400 upward muon events were collected over 5~years. This comparatively
low number, much lower than
what would correspond to the effective area,
reflects the notoriously large number of
failures of individual channels during each year. Still, NT200 could
compete with the much larger AMANDA for a while by searching for high energy
cascades {\it below} NT200, surveying a volume about ten times as large as NT200
itself \cite{Baikal-diff-Aynutdinov-2006}. In order to improve pattern
recognition for these studies, NT200 was fenced in 2005--2006 by three
sparsely instrumented outer strings (6 optical module pairs per string) which, however,
never delivered satisfying data.

\section{Detectors on the surface and underground}

The high threshold required to get a detector working in a hostile environment
such as the deep Pacific or the harsh conditions on the frozen Lake Baikal, resulted
in apparently long preparatory periods of both DUMAND and Baikal.
This led others to think about detectors near surface (for a review see \cite{shallow}). 
The advantages seemed tempting:
much easier access and a less challenging environment. Moreover,
proven techniques like tracking chambers or 
Cherenkov techniques $\grave{a}$ la Kamiokande could be used.  
However, with the exception of one, none of these projects was realized, be it by financial reasons, by 
the failure to convincingly demonstrate the background rejection
capabilities, or since shallow lake water parameters turned out to
be worse than expected. Also with the mentioned ``exceptional'' detector, the NEVOD tank in Moscow \cite{Nevod}, 
only a handful of low-significance neutrino candidates could be separated.

At the same time, underground detectors
moved from success to success. Remarkably, two of these successes had not been
on the top priority list of the experiments: neutrino oscillations (since the
trust in their existence was low in the eighties) and neutrinos from supernova
SN1987A (since Galactic or near-Galactic supernovae are rare).
At  the same time, neutrinos from high-energy astrophysics sources did not
show up. Even the data from largest detectors with about 1000\,m$^2$ area (MACRO
and Super-Kamio\-kande) did not show any indication of an excess over atmospheric
neutrinos. Seen from today, the search for
sources of high-energy neutrinos with detectors of 1000\,m$^2$ or less appears
to be hopeless from the beginning, with the possible exception of certain transient
Galactic sources. But when these detectors were constructed, this knowledge was 
not common and the search for point sources appeared as a legitimate (although not
priority) goal. 

\section{AMANDA}

In this situation, a new idea appeared on stage. 
In 1988,  Francis Halzen (University of Wisconsin, Madison) 
heard about a small test array of radio antennas at the Soviet Vostok station in Antarctica.
The Russians were going to test whether secondary
particles generated in neutrino interactions could be detected via their radio emission. 
Together with his colleagues Enrique Zas and Todor Stanev,
Halzen  realized that the threshold for this method was
discouraging high \cite{Halzen-icefishing}. Instead he asked himself whether the optical
detection via Cherenkov light, i.e. the DUMAND principle,
would be also feasible for ice. Halzen
remembers  \cite{Halzen-dreams}:
\textsl{\glqq I suspect that others must have contemplated the same
idea and given up on it. Had I not been completely ignorant about what was
then known about the optical properties of ice I would probably
have done the same. Instead, I sent off a flurry of E-mail messages to 
my friend John G.\,Learned,
then the spokesman of DUMAND. ... Learned immediately appreciated the 
advantages of an Antarctic neutrino detector.\grqq}

A few months later,
Halzen and Learned released a paper \textsc{High energy neutrino detection in deep
Polar ice} \cite{Halzen-1988}. With respect to the light attenuation length 
they \textsl{
``... proceeded on the hope that a simple test will confirm the belief
that is similar to the the observed 25\,m attenuation length for blue to mid UV
light in clear water in ocean basins.''}
Bubble-free ice was hoped to be found
at depths smaller than 1\,km. Holes drilled into the ice were supposed to refreeze
or, alternatively, to be filled with a non-freezing liquid. 

In 1989, two physicists of Buford Price's group at University of California, Berkeley,
tried to measure the ice transparency using existing boreholes at the South Pole. It would take, however,
another year until the first successful transparency measurement of natural
ice was performed  -- this time in Greenland. Bob Morse (Madison) and Tim Miller (Berkeley) lowered PMTs
into a 3\,km hole drilled by glaciologists \cite{Greenland-1990}.
In parallel to these first experimental steps, Price, Doug Lowder and Steve Barwick
(Berkeley), Morse and Halzen (Madison) and Alan Watson (Leeds)
decided to propose the Antarctic Muon and Neutrino Detection Array, AMANDA.

In 1991 and 1992, the embryonic AMANDA collaboration deployed PMTs
at various depth of the South Polar ice. Holes were drilled using a hot water
drilling technique which had been developed by glaciologists.  Judging the
count rate of coincidences between PMTs (which are due to
down-going muons), the light absorption length of the ice was estimated at
about 20\,m and scattering effects were supposed to be negligible.
 It should turn out later, that this was a fundamental misinterpretation
of the rates. But exactly this interpretation encouraged the AMANDA
physicists to go ahead with the project.

In 1992, the AMANDA collaboration was joined by the Swedish collaborators
led by Per-Olof Hulth (Stockholm). 
Steve Barwick (then already UC Irvine), designed a four-string detector
with a total of 80 PMTs. At Wisconsin, 
hot water drills were developed, in collaboration with the
Polar Ice Coring Office (PICO). AMANDA was located close to the
Amundsen-Scott station. Holes 
were drilled with pressurized hot water; strings with optical modules were deployed in the 
water which subsequently refreezes. 

This first AMANDA array was deployed in the austral summer 1993/94, at depths
between 800 and 1000\,m \cite{Amanda-shallow-1995}.  
Surprisingly, light pulses emitted at one string and registered at a neighbored string at 20\,m distance,
did not arrive after the expected 100\,ns, but were considerably delayed. 
The delay was due to light scattering at remnant bubbles. Light would not travel straight
but via random walk and become nearly isotropic after one or two meters,
a few times the distance called
effective scattering length which was 40\,(80)\,cm at 830\,(970)\,m depth, respectively. 
This made track reconstruction impossible \footnote{On the other hand the data proofed that
the {\it absorption} length was not 20\,m but of the order of 100\,m, i.e. much larger than in water.}

Nevertheless, our group from DESY joined. We were encouraged by the trend seen in the AMANDA data itself, as well as by ice core data taken at the Vostok station close to the geomagnetic South Pole: below 1300 meters bubbles should disappear. 
This expectation was confirmed with a second 4-string array which was deployed in
1995/96. The remaining scattering, averaged over 1500--2000\,m depth,
corresponds to an effective scattering length of about 20\,m. 
This was still considerably worse than for water but sufficient for track
reconstruction \cite{Amanda-ice-2006}.  Actually the angular resolution
for muon tracks turned out to be $2^\circ$--$2.5^\circ$ \cite{amanda-2004b}. 
Although better than for NT200 in Lake
Baikal ($3^\circ$--$4^\circ$), it was much worse than for ANTARES ($<0.5^\circ$,
see below). 
A big advantage compared to underwater detectors is the small photomultiplier
noise rate, about 0.5\,kHz in an 8-inch tube, compared to 20--40\,kHz due to
K$^{40}$ decays and bio-luminescence in lakes and oceans. 

\begin{figure}[ht]
\includegraphics[width=5.5cm]{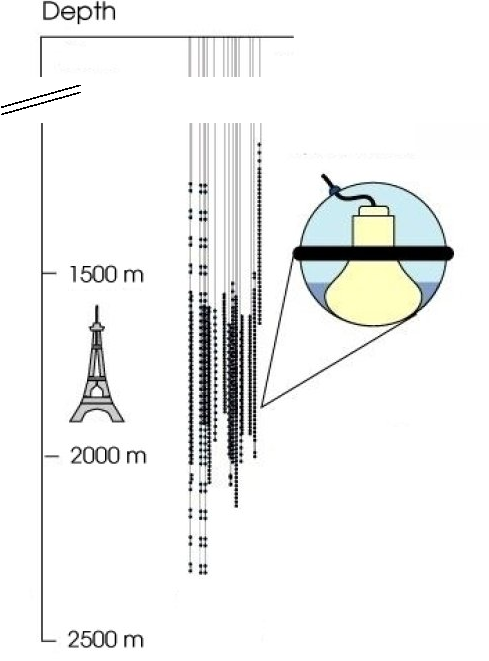}
\hspace{2.5cm}
\includegraphics[width=7cm]{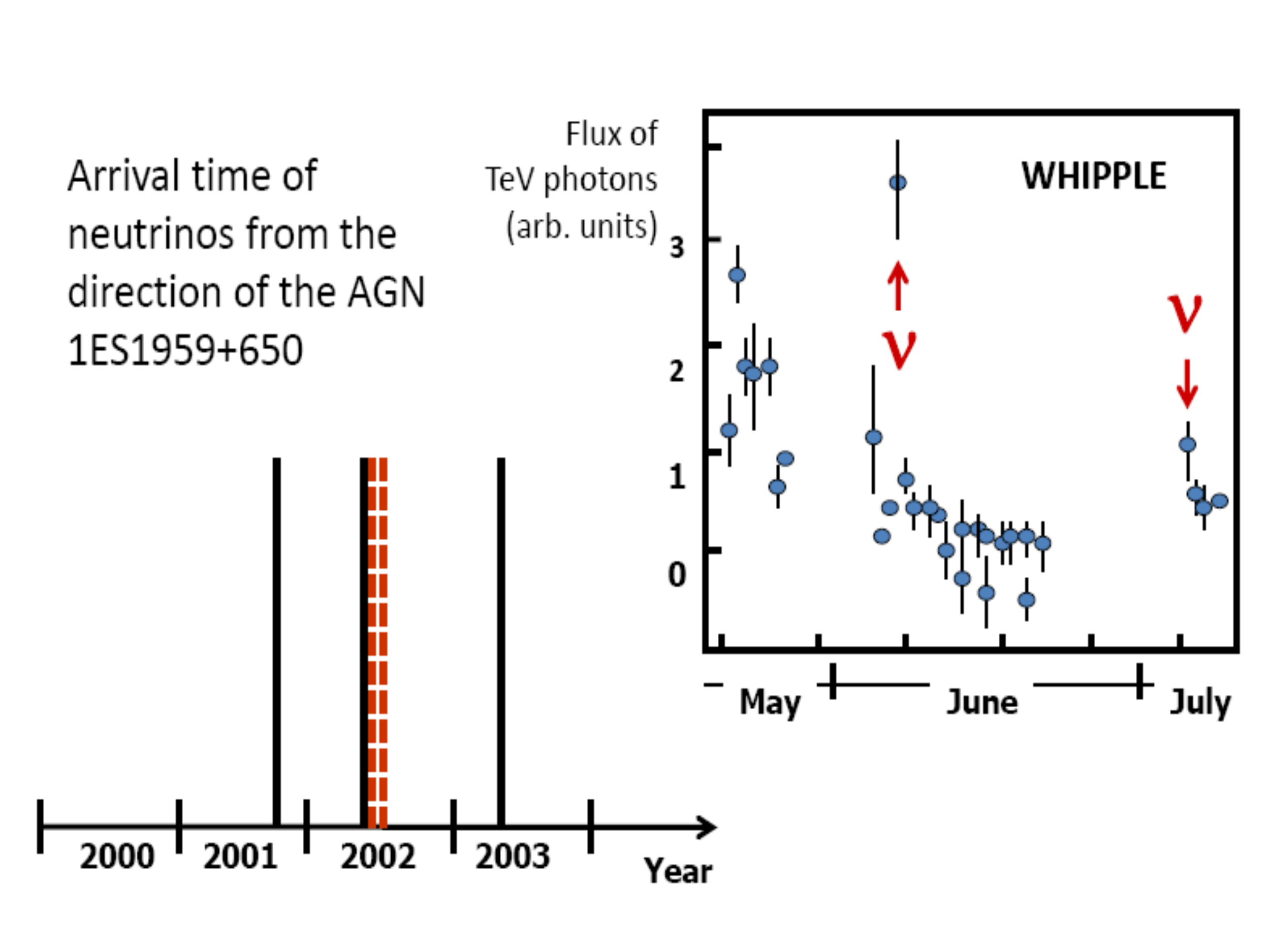}
\caption{    
{\it Left:} The AMANDA configuration. The detector consisted of 677 optical modules
at 19 strings. Three of the strings have been sparsely
equipped towards larger and smaller depth in order to explore ice properties,
one string got stuck during deployment at too shallow depth and was not used in
analyses. The Eiffel tower is shown to scale for size comparison.  {\it Right:} 
The "curious" coincidence of neutrino events from the direction of
an AGN with gamma flares from the same source. The second and third
of the three events recorded in 2002 (dashed) coincide within about one
day with peaks seen by the Whipple gamma-ray telescope.
\label{AMANDA}
}
\end{figure}

AMANDA was upgraded
stepwise until January 2000 and eventually comprised 19 strings with a total of
677 optical modules, most of them at depths between 1500 and 2000\,m.
Figure~\ref{AMANDA}, left, shows the AMANDA configuration. 

AMANDA was switched off in April 2009, after more than 9 years of data
taking in its full configuration.  The final neutrino sky map comprised
6959 events. 
No statistically significant local excess was observed, therefore 
only upper limits on point-source fluxes could be derived.
However, while analyzing
in 2005 the data taken from 2000-2003, five events where identified from the direction of
the Active Galaxy 1ES1959+650. Three of these came within 66~days 
in 2002 \cite{Markus-thesis}. Two of
the three neutrinos were coinciding within about a day with gamma-ray flares
observed by the gamma-ray telescopes HEGRA and Whipple
 -- see Fig.\ref{AMANDA}, right. One of these two flares was 
a so-called "orphan flare", i.e. it was not
accompanied by an X-ray flare, something one would
expect for a hadron flare.
This observation was quickly followed by two theoretical papers, one claiming
that the corresponding neutrino flux would not fit any reasonable assumption on
the energetics of the source \cite{Reimer-2005}, the other claiming that
scenarios yielding such fluxes were conceivable \cite{Halzen-Hooper-2005}. Since
the analysis was not fully blind, it turned out to be impossible to
determine chance probabilities for this event, and actually the result was never
published in a journal. However, it initiated considerations to send alerts to
gamma-ray telescopes in case time-clustered events from a certain direction would
appear. Such a "Target-of-Opportunity'' alert is currently operating
between IceCube and the gamma-ray telescopes MAGIC
(La Palma) and VERITAS (Arizona); a similar program is run by ANTARES.

Retrospectively AMANDA appears just as a prototype for IceCube. The final
goal of a cubic kilometer had been declared by Halzen before
AMANDA construction started \cite{Halzen-km3} and had actually been set by
the initial into-the-blue-designed dimensions of DUMAND. However, this does 
not reflect the {\it subjective} mid-1990s feelings/hopes of many experimentalists.
Figure \ref{Predictions}, right (taken from \cite{Hill-1996})
shows predictions for the number of events from a diffuse flux expected from all active galaxies for a variety of models. 
The five model versions with the highest fluxes
(although already at that time considered to be on the rather optimistic side) would have been easily detectable with DUMAND-II, AMANDA and ANTARES.
AMANDA, however, provided (only) record limits on fluxes for cosmic neutrinos, 
be it for diffuse fluxes, for point sources or for transient sources like Gamma Ray Bursts.
These limits ruled out the mentioned optimistic models on neutrino production in cosmic sources.
AMANDA also extended the measured spectrum of atmospheric neutrinos by nearly two orders of magnitude, from a few TeV to 200\,TeV. 
It also established record limits on indirect  dark matter search, on the
flux of magnetic monopoles, and on effects violating Lorentz invariance.
It would have detected neutrinos from a supernova burst in our Galaxy
(if such a burst would have appeared!), and it provided results on
the spectrum and composition of cosmic rays.

\vspace{-0.3cm}
\section{Flux predictions}
\vspace{-0.2cm}

\begin{figure}[ht]
\hspace{-2mm}
\includegraphics[width=8.7cm]{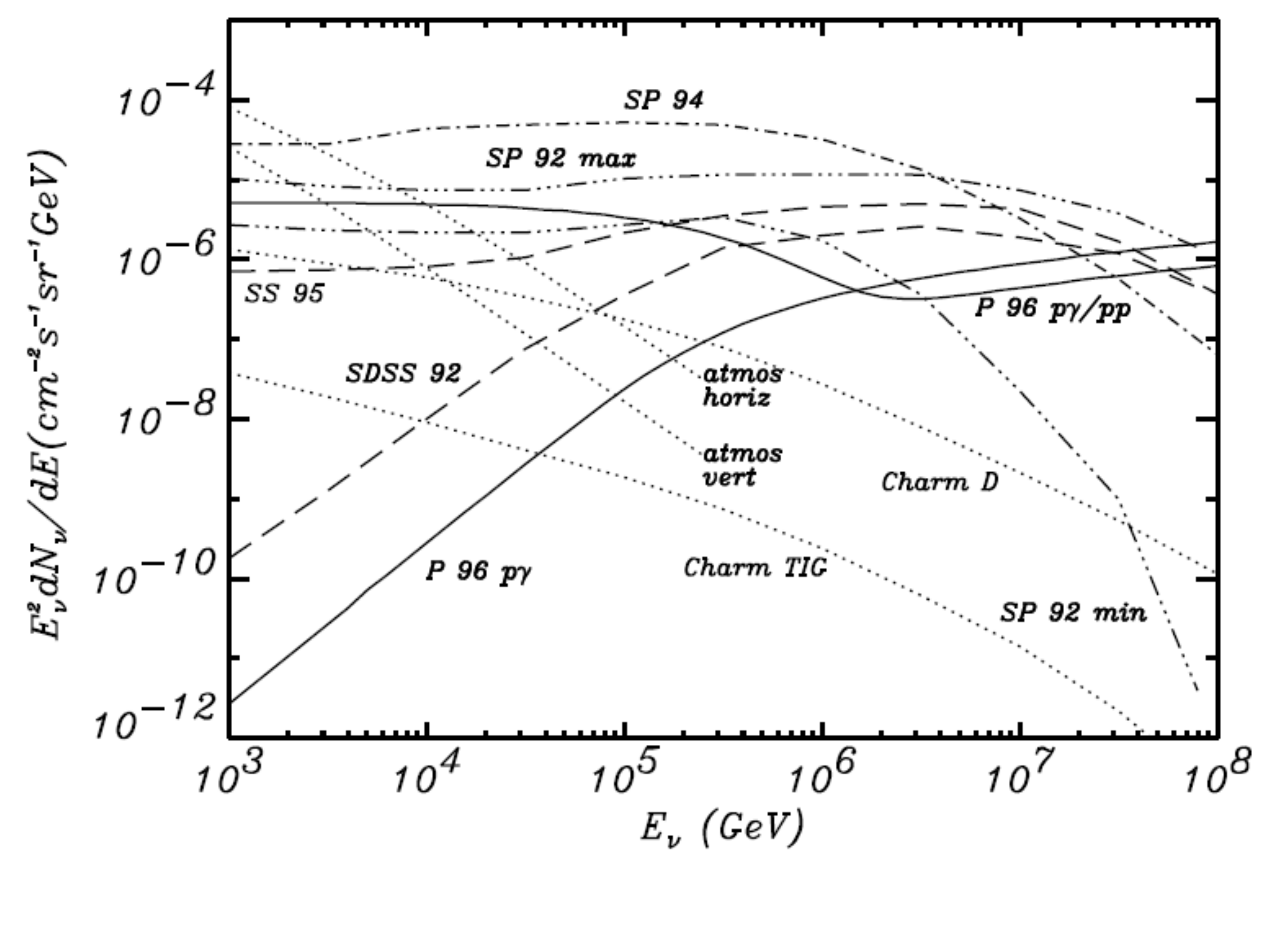}
\hspace{1mm}
\includegraphics[width=8.7cm]{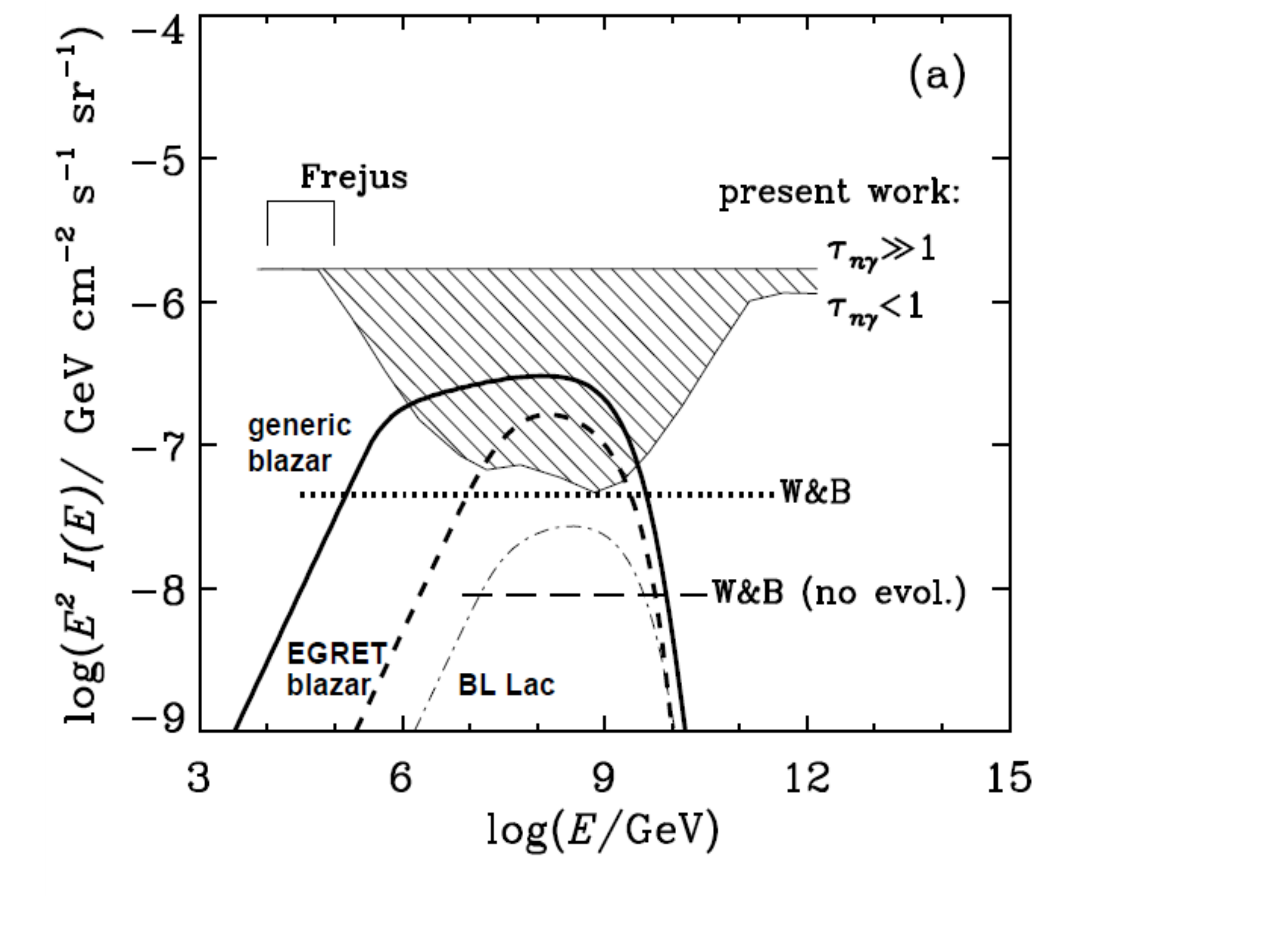}
\vspace{-1.0cm}
\caption{
Compilation of model predictions and bounds.  
{\it Left} (taken from~\protect\cite{Hill-1996}): Diffuse isotropic neutrino fluxes from the sum of all active galactic nuclei
predicted by various models (SDDS: Stecker,\,Done,\,Salomon and Sommers 1992, 
SP: Szabo and Protheroe 1994, SS: Stecker and Salomon 1992, P: Protheroe 1996 --
see~\protect\cite{Hill-1996} for references). Also shown are the horizontal and vertical flux of conventional atmospheric neutrinos from $\pi$ and K decays and the isotropic flux produced from prompt decays of charmed mesons in the atmosphere. 
{\it Right} (taken from~\protect\cite{Hettlage-Mannheim}): Cosmic ray bounds on extragalactic neutrino fluxes. The generic bound for the optically thin
($\tau_{n\gamma} < 1$) and thick case ($\tau_{n\gamma} \gg 1$),
the generic blazar bound (solid), the EGRET blazar bound, assuming that
for $L = 10^{48}$ erg/s the energy 
at which the neutron optical depth is unity has the value $E_{\tau=1} \approx 10^8$\,GeV
(dashed), and the BL Lac bound, assuming that 
$E_{\tau=1} \approx 10^{11}$\,GeV for $L = 3 \times 10^{44}$ erg/s 
(dot-dashed) are
shown together with the Frejus limit  and the bounds inferred by Waxman and Bahcall (W\&B)~\protect\cite{Waxman-Bahcall-1999} with and without
source evolution (see~\protect\cite{Hettlage-Mannheim} for references).
 }
\label{Predictions}
\end{figure}

Reported observations on TeV-PeV $\gamma$ rays from binary X-rays systems in the 1980s 
resulted in very high predictions for neutrino fluxes. However,  
more sensitive air shower arrays like CASA and Whipple did not confirm these observations, 
leading to 
upper limits on $\gamma$-ray fluxes which were 10-100 times lower than the original
detection claims, and consequently also to much lower neutrino fluxes. In a paper
from 1996, Gary Hill (at that time Adelaide) compiled the neutrino fluxes from models on the market and 
calculated the resulting event rates in a neutrino detector of 20,000 m$^2$ muon detection area
- that means for the scale of the AMANDA (and later ANTARES) \cite{Hill-1996}.
Fig.\ref{Predictions} gives the corresponding diffuse isotropic neutrino fluxes from the sum of all AGN as predicted by various models as well as the fluxes for neutrinos generated in the Earth's atmosphere. 
Predictions included models with charged pions generated in pp as well as in p$\gamma$ reactions.
The low-energy depletion for p$\gamma$ reactions is due to the energy threshold of protons if a pion would have to be produced off a photon target. Interestingly all pp models 
were in conflict by more than one order of magnitude with the so-called cascade bound 
which was derived already in 1975 by Berezinsky and Smirnov \cite{Berezinsky-1975}. This bound assumes that $\gamma$ rays co-produced with neutrinos -- if not visible at their original generation energy -- must have cascaded down and end up in a flux of gamma rays at lower energies.
  
The event numbers for a 20\,000 m$^2$ detector and muon energies larger than 10\,TeV, as given in \cite{Hill-1996}, are $\approx 40$ and 7 for the two considered p$\gamma$ models SDDS92 and P96 p$\gamma$, respectively. At least the flux for SDDS92 would have been easily detectable with AMANDA. 

In 1999, E.\,Waxman and J.\,Bahcall published their benchmarking paper on an upper bound for astrophysical neutrino fluxes 
\cite{Waxman-Bahcall-1999}. The authors did not start from $\gamma$-ray observations but
from the measured flux of charged cosmic rays at energies above $10^{17}$ eV. Assuming generic cosmic-ray
sources of one single type, they arrived at the bounds shown in Fig.\,\ref{Predictions},\,right. The two bounds are obtained, respectively, with and without a correction for neutrino energy loss due to redshift and for possible redshift evolution of the cosmic-ray generation rate. The value of the bounds is independent of energy. Shortly thereafter, K.\,Mannheim, R.\,Protheroe and J.\,Rachen~\cite{MPR} repeated these calculations but allowed the cosmic-ray flux to emerge from different source classes, each with a different cut-off but an envelope to all single spectra which fits the measured cosmic-ray spectrum. They confirmed the WB bound within a factor of 2, but only for
sources with a small optical depth for neutrons (i.e. neutrons could escape the acceleration region,
decay and constitute the observed proton spectrum) and for a limited energy range of 
$10^7$\,-\,$10^9$ GeV, see Fig.\,\ref{Predictions},\,right. For sources optically thick for neutrons, the
limit is constrained only by the extragalactic $\gamma$-ray background.

At that time, the only experimental limit on the diffuse neutrino flux was that of the 
Fr$\acute{e}$jus underground detector. It would be improved only in the early 2000s by data from AMANDA and NT200. 
The Fr$\acute{e}$jus limit is also shown in Fig.\,\ref{Predictions}, as well as two generic bounds on blazars.

It is clear from the above mentioned, that experimentalists in the late 1990s and early 2000s
had not yet given up the hope that AMANDA and ANTARES might detect a diffuse flux of
extragalactic neutrinos. Therefore the primary rational for building AMANDA and ANTARES was indeed to open a new neutrino window, rather than only being a prototype for IceCube and KM3NeT, respectively. 
Meanwhile, of course, we learned that one needs a cubic kilometer telescope to detect a diffuse extragalactic neutrino flux and that -- apart from well observable galactic point sources -- the detection of steady point sources may require even larger arrays.

\section{Neutrinos in the Mediterranean Sea}

With ongoing activities in Hawaii and at Lake Baikal and the first ideas on a telescope
in polar ice, the exploration of the Mediterranean Sea as a site for an underwater
neutrino telescope was natural. First site studies along a route
through the Mediterranean Sea were performed in 1989
by Igor Zheleznykh and his team of Russian physicists 
\cite{Deneyko-91}.
In July 1991, a Greek/Russian collaboration led by
Leonidas Resvanis from the University of Athens performed a cruise
and deployed a 
hexagonal structure carrying 10 PMTs
down to a depth of 4100\,m, close to Pylos/Peloponnesus. 
This was the start of the NESTOR project \cite{nestor-1994}.
 

In 1992, the collaboration included partners from Greece,
Germany, Russia, USA and Italy.
French institutes joined and left again to pursue their own
project ANTARES. More Italian institutes joined but later also 
decided to follow their own project, NEMO, close to Sicily.

NESTOR was conceived to consist of an array of hexagonal towers
covering an area of about $10^5$\,m$^2$.
A single tower should carry 168 PMTs on 12 floors \cite{nestor-1994}. 
After a long phase of tests and developments, 
a cable was installed to a site at 4\,km depth.
In 2004, a single prototype floor was deployed, connected and operated for
about one month \cite{nestor-test}. 
Then, its operation had to be terminated due to a failure of the cable to shore -- which was 
{\it de facto} the end of the project.

French collaborators temporarily had been members of NESTOR.
Led by Jean-Jacques Aubert (Marseille) and Luciano Moscoso (Saclay),
they pursued an independent strategy from the mid-nineties.  
In 1999, they presented  a full proposal for  
ANTARES (Astronomy with a Neutrino Telescope and Abyss environmental RESearch),
together with collaborators from Italy and the Netherlands \cite{antares-proposal}.
Currently the ANTARES collaboration includes in addition members from Germany, Morocco, Russia, Spain and Romania \cite{antares-web}.  

The construction of ANTARES started in 2002 with the deployment of a shore cable
and a junction box. The first string (or ``line'' in ANTARES nomenclature) was deployed in March 2006 and the last two of the 12 lines in May 2008. The years before 2007  were
extremely challenging learning years, but at the end successful. Similar to AMANDA after the air bubble effects at shallow depths, ANTARES' fate sometimes seemed hanging on a silk thread. 
But in 2008, the full detector was completed and is operational since then.

\begin{figure}[ht]
\includegraphics[width=7.5cm]{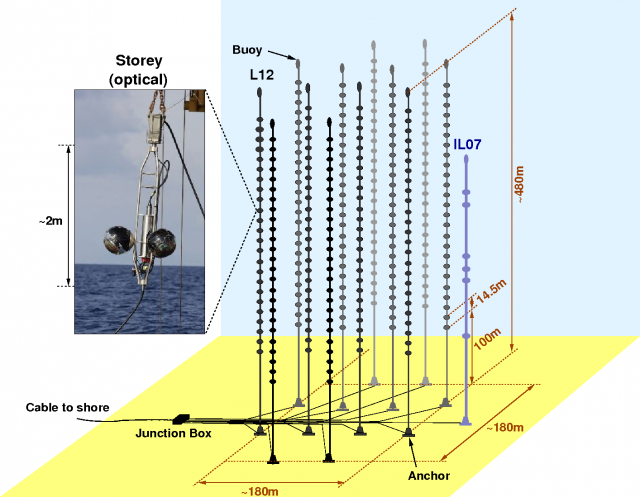}
\hspace{0.5cm}
\includegraphics[width=8cm]{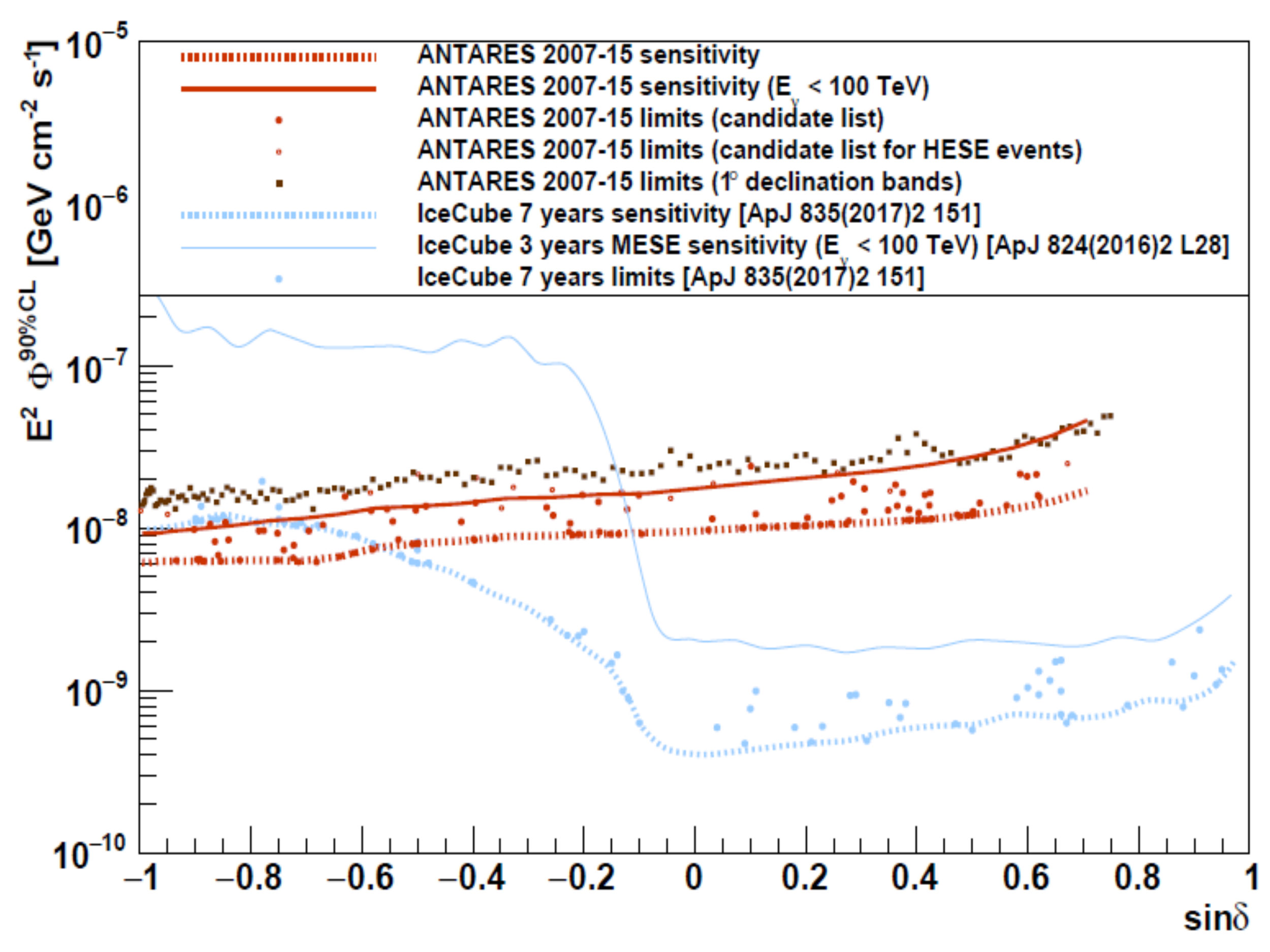}
\caption{
{\it Left:} Schematic view of the ANTARES detector. Indicated are the 12 strings and the instrumentation line. 
Shown as an inset is the photograph of a storey carrying 3 photomultipliers. {\it Right:}
Sensitivities and 90\% C.L upper limits on the signal flux  from ANTARES (8 years of data) and IceCube (7 and 3 years, respectively). Limits refer to selected source candidates. Curves are shown for an $E_{\nu}^{-2}$ spectrum without a cut-off and with a 
cut-off at 100\,TeV. See~\protect\cite{Antares-pointsources} for references and further explanations.}
\label{ANTARES}
\end{figure}

The lateral distance of the ANTARES strings is 60-70\,m. Each string carries 25 "storeys'' equipped
with three optical modules housing a 10-inch photomultiplier (Hamamatsu R7081-20).
The depth at the
ANTARES site is 2475\,m. The schematic setup is shown in Fig.~\ref{ANTARES} \cite{antares-detector}. A novelty of ANTARES was the all-data-to-shore concept conceived by
Maarten de Jong (Nikhef, Amsterdam). It avoided complicated off-shore hardware triggers and considerably simplified the full DAQ chain.

ANTARES has demonstrated that a stable operation of a deep-sea detector is possible.
Similar in size to AMANDA,  it has collected more than 7000 upward-going
muon tracks over eight years of operation.
With its excellent view of the Galactic plane and
good angular resolution, the telescope could constrain the Galactic origin of the 
cosmic neutrino flux reported by IceCube. ANTARES has explored the Southern sky and in particular
central regions of our Galaxy searching for point-like objects,
for extended regions of emission, and for signals from transient objects selected
through multi-messenger observations.
The good angular resolution for cascades
allowed to include them in point-source searches 
\cite{Antares-pointsources}. Fig.\,\ref{ANTARES} shows the limits obtained from eight years of ANTARES data compared to the limits from the much larger IceCube detector (see the following sections). It demonstrates, for a spectrum with a 100 TeV cutoff, how IceCube becomes almost blind  for neutrinos from the South, and for an uncut spectrum, how far into the Southern hemisphere IceCube and ANTARES data can be profitably added. Actually, the recent two years resulted in a number of joint ANTARES/IceCube analyses on Galactic neutrino point sources, neutrinos from the direction of gravitational wave events and for neutrinos from Dark Matter annihilations in the Galactic plane or the Galactic center (see section \ref{sec-status}).

The third Mediterranean project was NEMO, 
(NEutrino Mediterranean Observatory) \cite{nemo-2009},
launched in 1998 after Italian groups left NESTOR. 
Leading persons of this initiative were Antonino Capone (Rome) and
Emilio Migneco (Catania).
Rather than building a separate detector of medium size, their objective was to study the
feasibility of a cubic kilometer detector, to develop corresponding
technologies and to identify and
explore a suitable site, in their case close to Sicily.
The basic unit of NEMO are towers composed by a sequence of 
floors. Different to NESTOR, floors consist of horizontal "bars", 
originally foreseen to be 15 m long and
each equipped with four 10-inch PMTs.
A suitable site was found at a depth of
3.5\,km, about 100\,km off Capo Passero on the South-Eastern coast of Sicily. 
Actually this will be the location of two of the KM3NeT/ARCA blocks (see below). A variety of NEMO tower versions have been deployed over the years, and eight of such
towers will likely be linked to the ARCA blocks.

\section{IceCube}
\label{sec-icecube}

With IceCube \cite{icecube-web}, the idea of a cubic-kilometer detector 
was finally 
realized. 
Actually, the first initiative beyond AMANDA was a concept called {\it DeepIce},
a proposal for multidisciplinary investigations, including neutrino and
cosmic ray astrophysics, glaciology, glacial biology, seismology and climate research.
DeepIce was proposed in 1999 to NSF, but was not funded. 
As a consequence, already in November of the same year
a first 67-page IceCube proposal was submitted to NSF. 
It was signed essentially by the collaborators of the old AMANDA collaboration.
Soon, a number of additional institutions became interested and a new collaboration
was formed, the IceCube collaboration, which meanwhile has grown to more
than 50 institutions from 12 countries.
Paradoxically, the two collaborations co-existed until 2005, then joining
to one collaboration, IceCube.

A first, single string of IceCube was
deployed in January 2005. The following seasons resulted
in 8,\,13,\,18,\,19,\,20 and 7 strings, respectively. 
The last of 86 strings was deployed on Dec.\,18,\,2010.


\begin{figure}[ht]
\includegraphics[width=6.5cm] {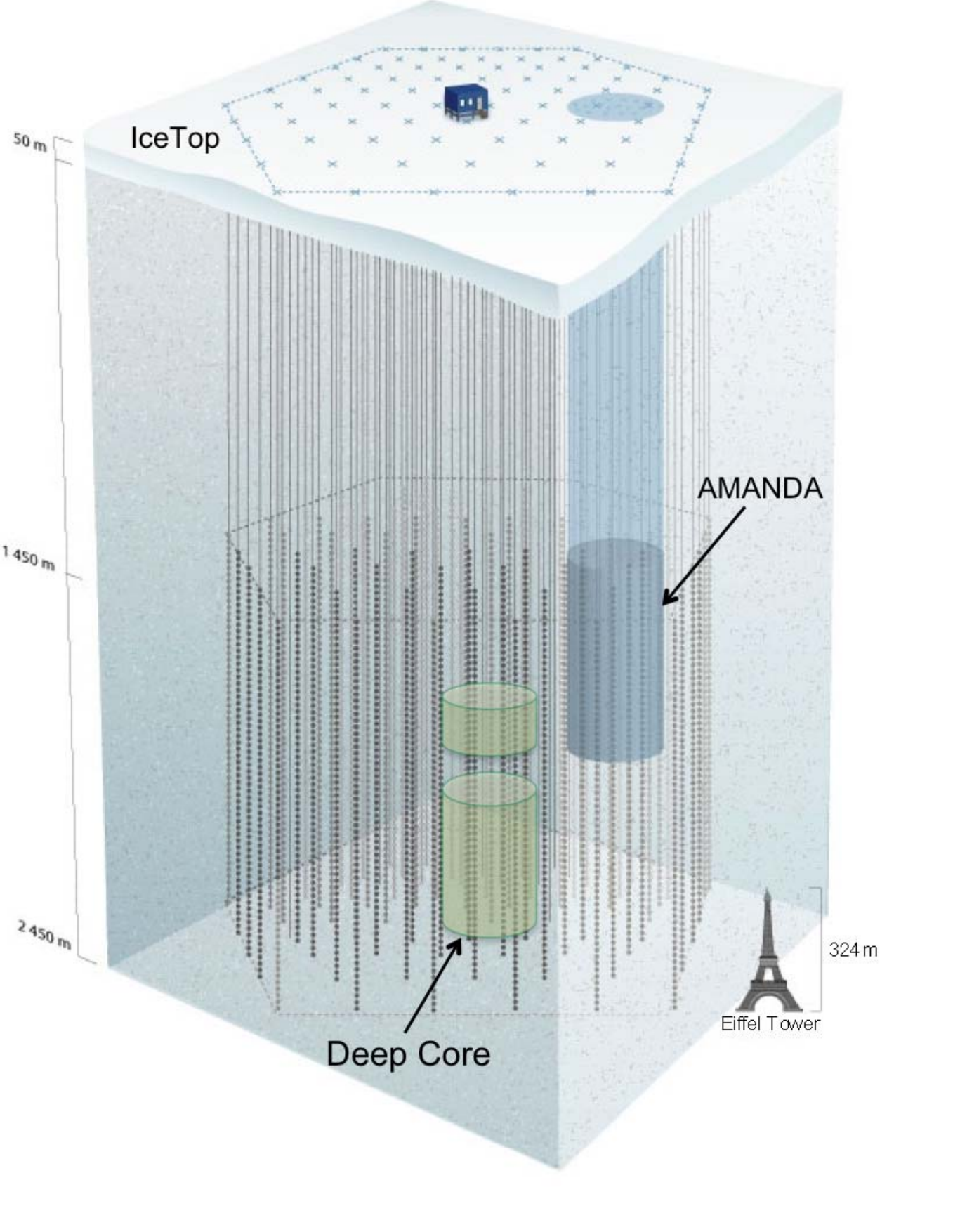}
\includegraphics[width=10.5cm] {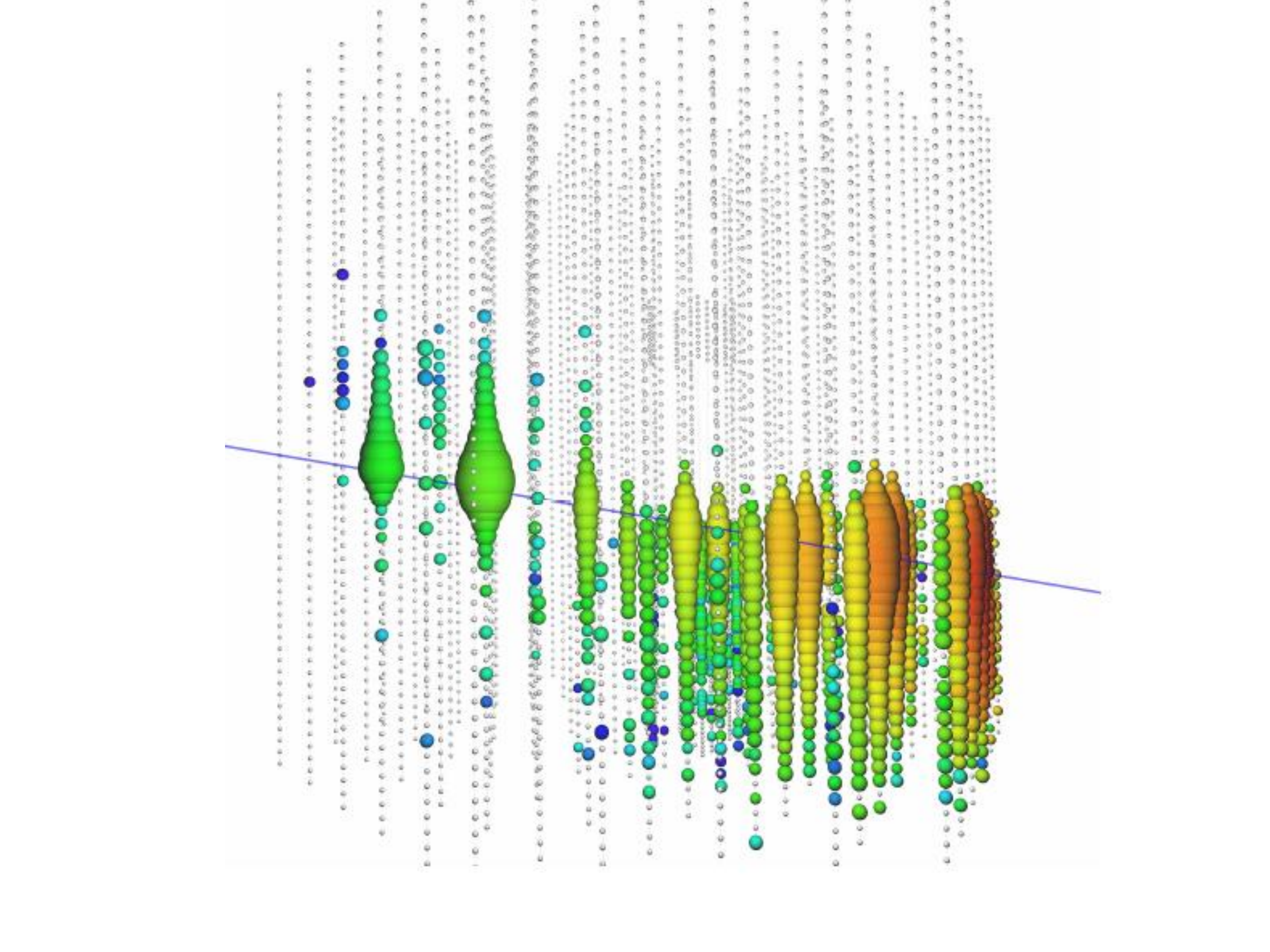}
\caption{
{\it Left:} Schematic view of the IceCube neutrino observatory. AMANDA was replaced by
DeepCore, a nested low-threshold array. At the surface, the air shower array
IceTop and the IceCube counting house are indicated. {\it Right:} The upward moving muon track 
with the highest deposited energy, $2.6\pm0.3$\,PeV. Taking into account
the muon energy loss between generation point and IceCube and the energy transfer
from the parent neutrino to the muon, the most probable energy of the parent neutrino 
turns out to be in the range of 7 to 9 PeV.}
\label{IceCube}
\end{figure}

IceCube consists of 5160 digital optical modules (DOMs) installed on 86 strings at depths of 1450 to 2450\,m.
A string carries 60~DOMs with 10-inch photomultipliers Hamamatsu 
R7081-02 housed in a 13-inch glass sphere. 
Signals  are digitized in the DOM and sent to shore via copper cables.
320~further DOMs are installed in IceTop, an array of detector stations on the
ice surface directly above the strings (see Fig.~\ref{IceCube}). AMANDA,
initially running as a low-energy sub-detector of IceCube, was decommissioned in
2009 and replaced by DeepCore, a high-density sub-array of six strings at large
depths (i.e.\,in the best ice layer) at the center of IceCube. 
DeepCore collects photons with about six times the efficiency of full
IceCube, due to its smaller spacing, the better ice quality and the
higher quantum efficiency of new PMTs.
Together with the veto provided by IceCube, this results in an expected
threshold of about 10\,GeV. This opened a new window for oscillation physics
and indirect dark matter search.

The muon angular resolution achieved by present reconstruction algorithms is
about $1^\circ$ for 1\,TeV muons and below $0.5^\circ$ for energies above 10 TeV. 
Much more effectively than underwater detectors (with their environment of high optical noise),
IceCube can be operated in a special mode to detect burst neutrinos from supernovae. The low dark-count rate of the PMTs  allows for detection of the
feeble increase of the summed count rates of all PMTs during several
seconds, which would be produced by millions of interactions of few-MeV
neutrinos from a supernova burst \cite{icecube-sn-2011}. 

\vspace{-0.3cm}
\section{Where do we stand?}
\label{sec-status}
\vspace{-0.2cm}

\begin{figure}[h]
\vspace{-0.1cm}
\centering
\includegraphics[width=11cm]{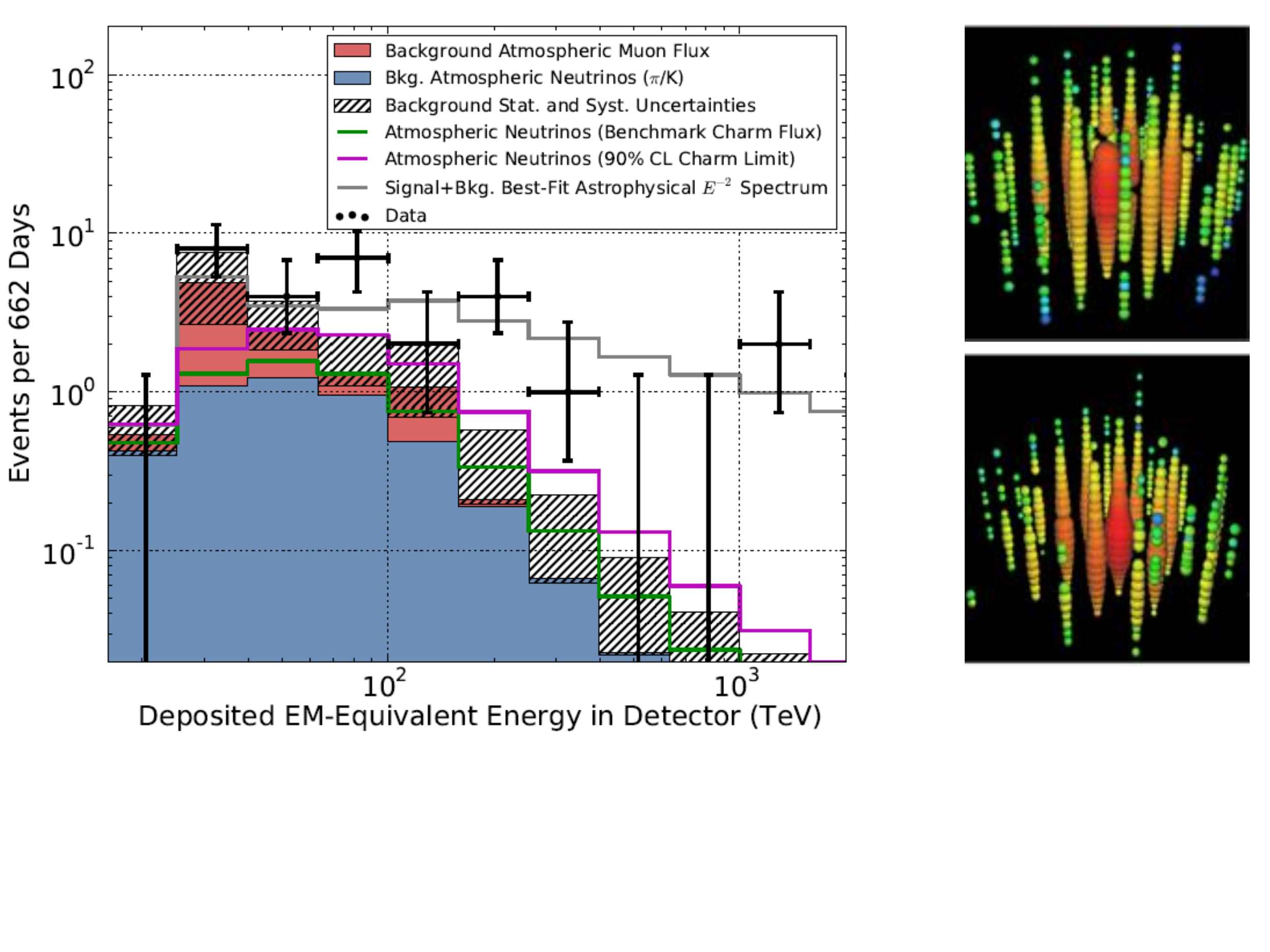}
\vspace{-2cm}
  \caption{
{\it Left:} Distribution of the energy deposited by 27 events from the two-year HESE analysis. 
Backgrounds of atmospheric origin 
come from punch-through down-going muons and from atmospheric neutrinos. 
While the flux of neutrinos from $\pi$ and K decays is well known (blue region), the neutrino flux 
from charm decays in the atmosphere is uncertain and dominates
the uncertainty of all background sources (gray region with 1$\sigma$ uncertainties). The best-fit with an 
$E_{\nu}^{-2}$ astrophysical spectrum is shown as gray line~\protect\cite{Science-2013}. 
{\it Right:} Event displays of the two events dubbed ``Bert'' (top, August 2011, 1.04 PeV) and ``Ernie'' (bottom, January 2012, 1.14 PeV).
}
\label{Ernie-and-Bert}
\end{figure}

It has been predicted since long that the first evidence for extragalactic cosmic neutrinos would be provided by a diffuse flux rather than by single-source signals 
\cite{Lipari}. 
The first tantalizing hint to cosmic neutrinos in IceCube came from two shower-like events with energies 
$\approx 1$ PeV, discovered in 2012 and dubbed ``Ernie'' and ``Bert'' 
(see Fig.\ref{Ernie-and-Bert}, right).
A follow-up search using the same data (May 2010 to April 2012) with a lowered threshold (30 TeV) 
provided 25 additional events. This analysis used only events starting in a fiducial volume of 
about 0.4 km$^3$ (High Energy Starting Events, or ``HESE''), using the other 60\% of IceCube as veto against all sorts of background. Energy spectrum and zenith angle distribution
of the 27 events excluded an only-atmospheric origin with $4.1\sigma$ and suggested that 
about 60\% were of cosmic origin, 
at energies above 100 TeV even about 80\% \cite{Science-2013}.
Figure\,\ref{Ernie-and-Bert},\,right shows the  energies deposited by these events inside IceCube. With increased statistics, the excess has meanwhile reached a significance of more than $7\sigma$. These analyses can be considered a breakthrough, and it was observed in the diffuse flux, as suggested in \cite{Lipari}.

A $>5\sigma$ excess of high-energy cosmic neutrinos is also seen in the spectrum of secondary muons generated by neutrinos that have traversed the Earth, with a zenith angle less than 5 degrees above the horizon
(``upward through-going muons'') \cite{muons-7years}. It is this sample to which the muonic event with the highest energy deposition shown in Fig.\ref{IceCube} belongs.

While both analyses (HESE and through-going muons)  have reached a significance for a strong non-atmospheric contribution of more than 5$\sigma$, the spectral indices 
of the astrophysical flux from both analyses are in tension, with a harder spectrum
for the muon sample. To what an extent this difference emerges from the slightly different
energy ranges of the two analyses, or from systematic effects related to the ice medium, has
still to be resolved and requires independent scrutinisation by water neutrino telescopes of the cubic kilometer scale.  

\begin{figure}[ht]
\hspace{-1cm}
\includegraphics[height=6cm]{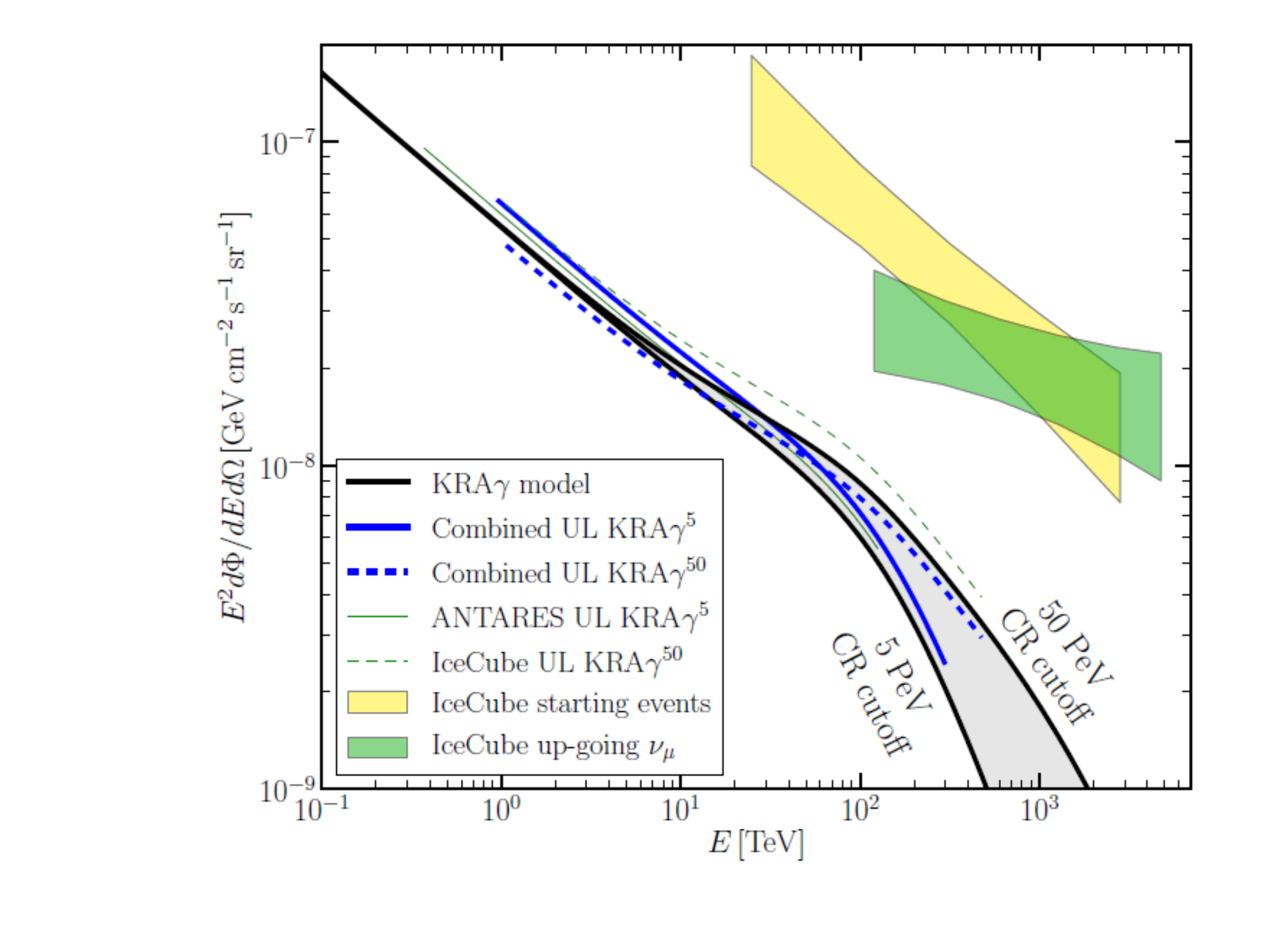}
\hspace{0.5cm}
\includegraphics[height=6cm]{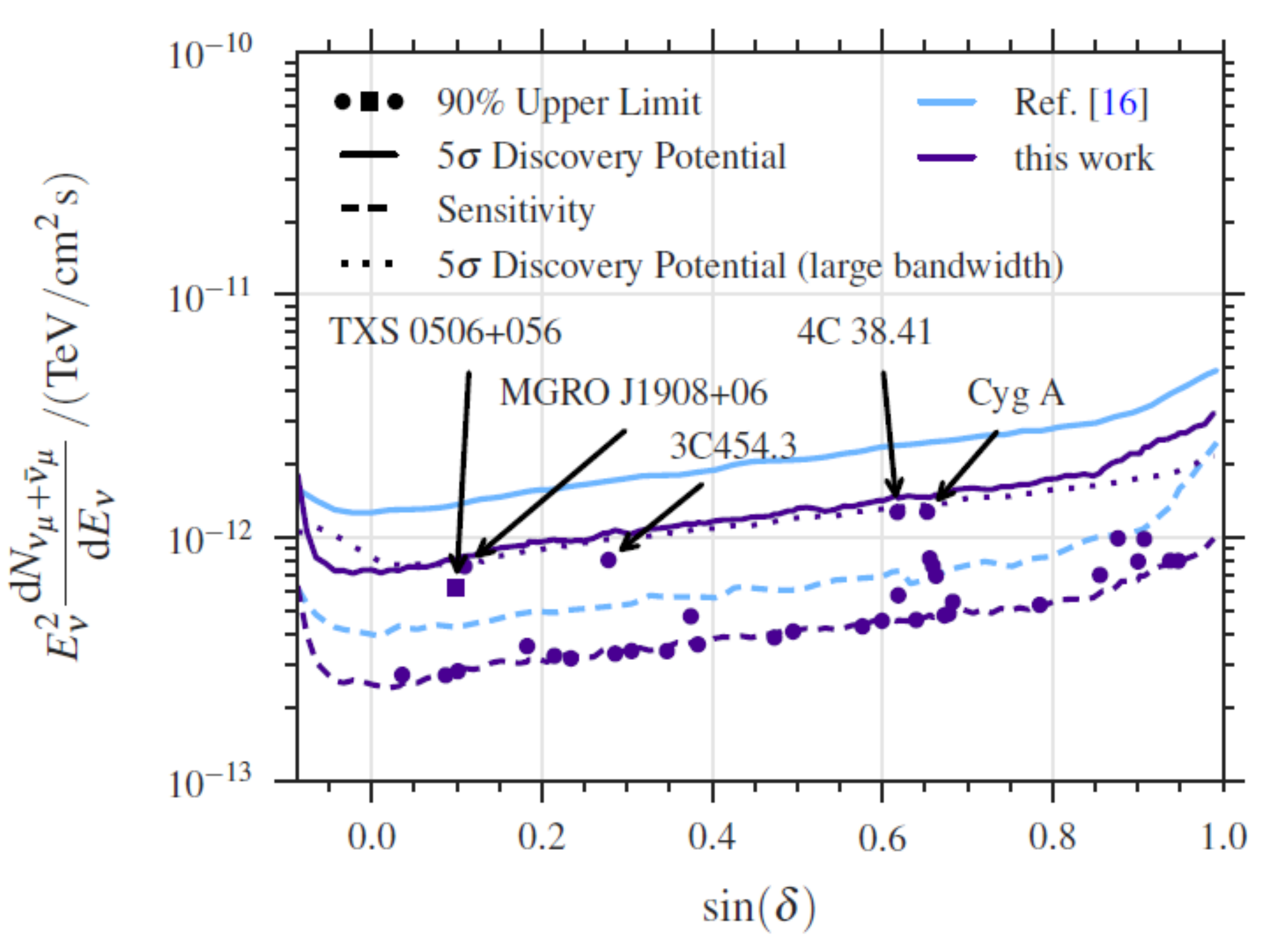}
\vspace{-4mm} 
\caption{
{\it Left:}
Combined upper limits at 90\% confidence
level (blue lines) on the three-flavor neutrino flux of the
KRA model~\protect\cite{KRA} with 5 and 50 PeV cutoffs (black lines).
The boxes represent the diffuse astrophysical neutrino 
fluxes measured by IceCube using an isotropic 
flux template with HESE events (yellow) and upgoing tracks (green)~\protect\cite{Galactic-Plane-combined}.
{\it Right:} Sensitivity (dashed) and 5$\sigma$ discovery potential (solid) on the
flux normalization for an $E^{-2}$ source spectrum as function of the declination~\protect\cite{IC-pt-8years}.
Results are for 8 years data. The progress due to better analysis methods and more statistics is
demonstrated by comparison to the 7-year results (blue lines~\protect\cite{IC-pt-7years})
90\% C.L. Neyman
upper limits on the flux normalization for sources in an {\it a priori}
and monitored source list are shown as circles and squares, respectively.
}
\label{GP-pts}
\end{figure}

Naturally the question arose, how much of this diffuse flux can be related to our Galaxy. In the latest publication on this issue, ANTARES and IceCube jointly analysed their data to identify an excess from the Galactic Plane \cite{Galactic-Plane-combined}. Such an excess is expected not only from individual sources but dominantly from cosmic ray interactions with Galactic gas and radiation fields.The resulting limits are shown in 
Fig.\,\ref{GP-pts},\,left and are compared to the  flux of the HESE and highest-energy $\nu_{\mu}$ data. They exclude that more than 8.5\% of the observed diffuse astrophysical flux come from the Galactic plane. This limit is already substantially constraining model predictions  \cite{KRA} (gray band). Note that for this analysis, the contribution from ANTARES dominates due to their better visibility of the central parts of the Galaxy. 

Also the search for steady point sources remained unsuccessful until now. 
Figure \ref{GP-pts},\,right shows sensitivities and upper limits from an IceCube  Northern hemisphere search, using eight years of data with 497\,000 upward muons. The impressive factor-2
improvement of the sensitivity by adding only one year is mainly due to improved reconstruction and analysis methods. From the year-2000 point source limits published with AMANDA 
\cite{Amanda-B10} to these
limits, the sensitivity has improved by a factor of $\approx$\,2000.

From several model predictions one could conclude that an improved angular reconstruction and twice more data could bring us close to discovery. For blazars, however, this hope is downsized by various 
blazar stacking analyses, none of them yielding an excess in the directions of blazars. 
The most recent IceCube stacking analysis \cite{blazarstacking} indicates that only 4-6\% of the observed diffuse astrophysical muon neutrino flux could come from typical blazars.

To improve the signal-to-background ratio one can search for transient signals, preferentially
in coincidence with an observation in electromagnetic waves. Examples are flares of Active Galactic Nuclei (AGN) or Gamma-Ray Bursts (GRB). Some models
assume GRBs are the dominant source of
the measured cosmic-ray flux at highest energies, either by neutron escape \cite{Ahlers}
or by escape of both neutrons and protons \cite{WB-GRB} from the relativistic fireball. 
All three collaborations -- Baikal, ANTARES and IceCube -- have searched for neutrinos in local and spatial coincidence with GRBs. The most recent IceCube limits \cite{IC-GRB} exclude
even the more conservative of the two predictions \cite{WB-GRB} at more than
90\% confidence level, greatly constraining the hypothesis that GRBs
are significant producers of ultra-high energy cosmic rays in the prompt GRB phase.

The search for transient signals is also the rationale for follow-up observations in electromagnetic waves which are triggered by neutrino observations in IceCube or ANTARES. 
Eventually, a real-time alert issued by IceCube on September 22, 2017, led to the
first coincident observation of a high-energy energy neutrino 
(290 TeV estimated neutrino energy) with electromagnetic information. Fermi and MAGIC
were the first to confirm that a blazar named TXS 0506+056 was the likely source of the neutrino and that
it was in an active state at the time of the alert \cite{TXS1}. IceCube examined its archival data in the direction of TXS 0606+056 and found an additional 3.5$\sigma$ evidence for a flare of thirteen neutrinos starting at the end of 2014 and lasting about four months \cite{TXS2}. These observations
where communicated as ``compelling evidence'', but certainly they
do not mark yet a clear discovery. However, the next outbreak of a galaxy may be observed very soon -- and possibly recorded not only by
IceCube but also by neutrino telescopes just under construction, like KM3NeT in the
Mediterranean Sea and GVD in Lake Baikal. Then, September 22, 2017 will likely be counted as
the exciting start for a precise mapping of the neutrino sky.

I want to close this section with the display of two events which particularly impressively illustrate the long way its takes from an idea to observation. 

35 years ago, V.\,Berezinsky and A.\,Gazizov suggested \cite{Glashow-Berezinsky}, that the 
``Glashow resonance'' \cite{Glashow} $\bar{\nu_e}e\rightarrow W$  would appear at 
$E_{\bar{\nu_e}} = 6.3$ PeV for a neutrino interaction with an electron resting in the lab system . This is the only uncontroversial way to separate anti-neutrinos from neutrinos in large underwater detectors. The typical visible energy in such an event would be 
somewhat below 6\,PeV.
It was quite recently, that a first candidate for such an event has been observed \cite{Glashow-IceCube}, see Fig.\,10,\,left.
The event is partially-contained, with a deposited energy
of $5.9\pm 0.18$\,PeV and with a possible hadronic character of the cascade. 
This (still preliminary) result may define the final pillar of an impressing bridge, with three and a half decade between the prediction and the observation of this very special signature!

\vspace{2mm}

\begin{figure}[ht]
\hspace{0.5cm}
\includegraphics[height=6cm]{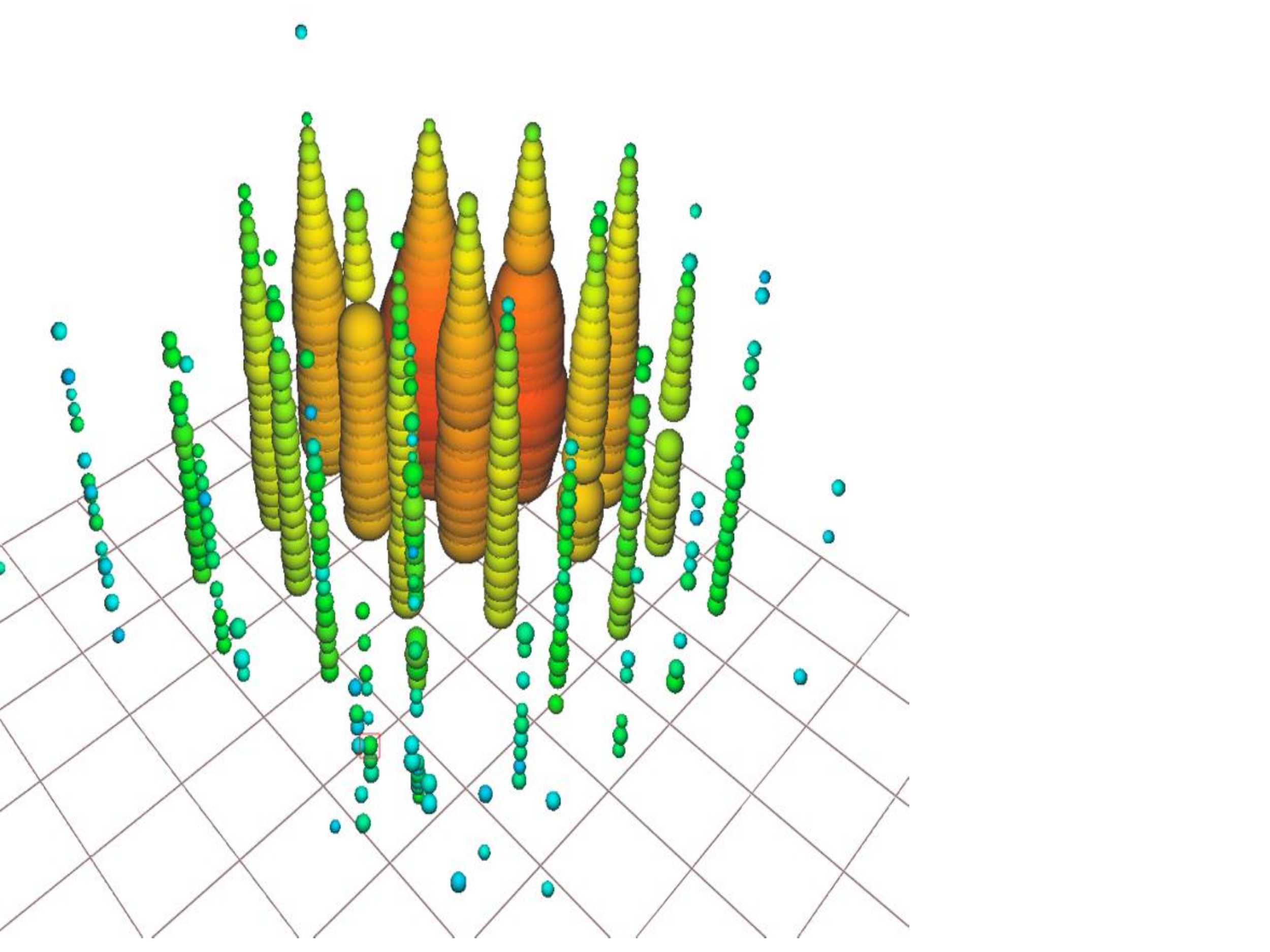}
\hspace{0.5cm}
\includegraphics[height=6cm]{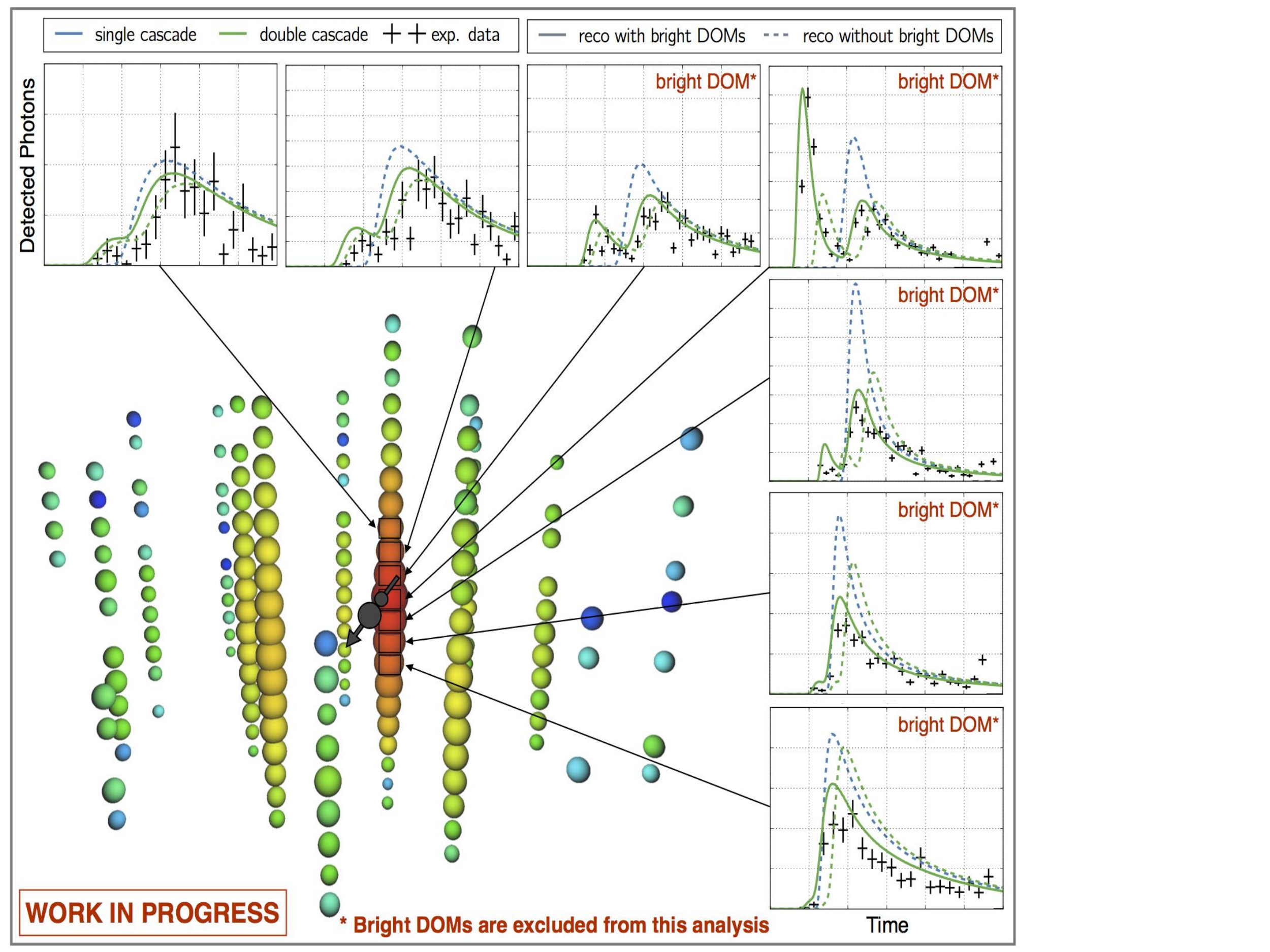}
\caption{{\it Left:} IceCube's first candidate for a Glashow event, with a reconstructed energy of $5.9\pm 0.18$\,PeV (statistical errors only). {\it Right:} The second of IceCube's double-cascade candidates. The reconstructed cascade positions are indicated as gray circles, the direction with a gray arrow. The framing diagrams show the double-pulse form detected in some of the optical modules. Both results are still preliminary (status 2018).
	}
\label{Glashow-DoubleBang}
\end{figure}

The idea for the second event goes back to 1994, when J.\,Learned and S.\,Pakvasa suggested that an underwater neutrino detector could prove the existence of the $\nu_{\tau}$ and measure its mixing with other 
flavors \cite{doublebang-Learned}. Charged current $\nu_{\tau}$ events would produce a
double cascade. At a few PeV the two ``bangs'' (one for the primary interaction, the other from the $\tau$ decay) would be separated by roughly 100\,m
and connected only by the minimum ionizing $\tau$ track. This is the possible signature of two events found in the data samples of 2012 and 2014. Figure\,10,\,right
shows the second of these events. The reconstructed energies of the first and the second cascade are 9\,TeV and 80\,TeV, respectively, and their distance is 17 meters \cite{Juliana}.

\section{Concluding remarks}

From the first ideas on detecting high-energy cosmic neutrinos to their actual detection it took slightly more than half a century, from 1960 to 2013. The technological challenges turned out to be tremendous. From the first ideas to build a cubic kilometer detector (DUMAND, 1973-78) to the completion of IceCube in 2010 it took about 35 years. The first 
atmospheric underwater/ice neutrinos have been identified in Lake Baikal and AMANDA in the mid 1990s. The first literal successor of DUMAND, the {\it deep-sea} detector ANTARES, was completed in 2008. Both
ANTARES and IceCube are presently running and take high-quality data.

The flux predictions over that period went from ``unknown'' to ``high'' and then, steadily, to something which one could call ``desperately low''. 
Five years after the detection of cosmic neutrinos, we have learned a lot about their
spectrum and flavor composition. 
We have learned that typical blazar jets and GRBs can contribute only a small fraction to
the observed astrophysical neutrino flux (which does not exclude that a sub-class of blazars and GRBs, even not labeled as such, might do the job).
No individual steady sources have been detected yet. The non-observation of neutrinos coinciding with GRBs strongly constrains models which attribute the highest-energy cosmic rays to GRBs. Instead, evidence for a possible first transient neutrino source, the blazar 
TXS 0506+056 has been obtained.
No neutrinos have been observed that could be attributed to the GZK effect \cite{BZ}, but the non-observation starts constraining evolution scenarios
for ultra-high energy cosmic rays sources. The detection of neutrinos from GZK interaction will likely have to wait for much larger detectors which use radio or acoustic detection technologies (which are not addressed in this review).

The next important steps are being done at the Northern hemisphere: The Gigaton Volume Detector in Lake Baikal (Baikal-GVD) \cite{GVD} and KM3NeT in the Mediterranean Sea \cite{KM3NeT}. At the time of writing this paper, Baikal-GVD comprises three of the eight clusters of its 
Phase\,1. This phase with about 0.4\,km$^3$ effective volume is planned to be completed in 
2021. KM3NeT is developing along two paths, its high-energy incarnation ARCA, close to Sicily, and the low-energy array ORCA close to Toulon. ARCA is primarily devoted to high-energy neutrino astromomy, ORCA to oscillation physics. In a first phase, ARCA is planned to consist of two blocks of 0.6\,km$^3$. At the same time the IceCube collaboration, on its way towards a 6-10\,km$^3$ 2$^{nd}$ generation detector (IceCube-Gen2) is working on a moderate upgrade (IceCube-upgrade) which will help a better understanding of the existing detector and allow testing technologies for IceCube-Gen2. In less than a decade we very likely will have a few kilometers of instrumented water/ice both at the Northern and the Southern hemisphere.  Under the umbrella of the Global Neutrino Network GNN \cite{GNN}, the various collaborations are presently closely cooperating, and these links will tighten in the future. 

We have made more than a factor-of-thousand step in sensitivity compared to the beginning of the century.
This is far more than the traditional factor of ten which so often led to the discovery
of new phenomena \cite{Harwit}. In our case, the first firm discovery has been made with the detection of a diffuse cosmic neutrino flux. I am convinced that the next steps will deliver the long-thought individual sources and allow starting to map the high-energy neutrino sky. But this is future, not history ...

\noindent
{\bf Acknowledgements} I thank Michel Cribier, Jacques Dumarchez and Daniel Vignaud for organizing this impressive conference and J\"urgen Brunner, Francis Halzen and Uli Katz for helpful remarks on the manuscript.

\end{document}